\documentclass[preprint]{aastex}
\usepackage{graphicx}

\def\etal{{\it et al.}~}

\begin{document}

\title{Planetesimal disk evolution driven by 
planetesimal-planetesimal  gravitational scattering.}

\author{R. R. Rafikov}
\affil{Princeton University Observatory, Princeton, NJ 08544}
\email{rrr@astro.princeton.edu}

\begin{abstract}
We investigate the process of an inhomogeneous 
planetesimal disk evolution caused by the 
planetesimal-planetesimal  gravitational scattering. We develop a rather
general approach based on the kinetic theory  
which self-consistently describes the evolution in time 
and space of both the disk surface density and 
its kinematic properties --- dispersions of eccentricity and inclination. 
The gravitational scattering of planetesimals is assumed to be in the 
dispersion-dominated regime which considerably simplifies analytical 
treatment. The resultant equations are of advection-diffusion type.
Distance dependent scattering coefficients entering these equations 
are calculated analytically under the assumption of two-body scattering 
in the leading order in Coulomb logarithm. They are essentially nonlocal 
in nature. Our approach allows one to explore the 
dynamics of nonuniform planetesimal disks with arbitrary mass 
and random velocity distributions. It can also naturally include other
physical mechanisms which are important for the evolution of such
disks --- gas drag, migration, and so on.
\end{abstract}



\section{Introduction.}  
\label{sect:intro}


The discovery of extrasolar giant planets around nearby stars 
(Mayor \& Queloz 1995; Marcy \etal 2000; Vogt \etal 2000; 
Butler \etal 2001) 
has been one of the most exciting astrophysical findings of 
the last decade. It has revived interest in planetary 
sciences and stimulated many theoretical studies.
One of the areas which has received a lot of attention recently is
the formation of terrestrial planets --- planets like Earth or Venus.

There are many reasons for this interest. First of all, our own Solar System
hosts several terrestrial planets and we must understand their
formation mechanisms if we want to know the history of our planetary 
system. Second, despite the huge differences in their physical properties,
giant planets are probably linked to the terrestrial ones through 
their formation mechanism, since it is widely believed that giant planets
form as a result of gas accretion on a preexisting rocky core which 
essentially was a massive Earth-type planet (Mizuno 1980; Stevenson 1982; 
Bodenheimer \& Pollack 1986; see however
Boss \etal 2002). Thus terrestrial planets are probably an important
component in the evolutionary history of the giant planets.
Third, a lot of effort is currently being directed toward
 designing and building
space missions with the ultimate goal of  
detecting Earth-type planets around other stars. 
Theoretical understanding of how terrestrial planets had come into being 
will help us plan these missions most effectively. 
Finally, observations of IR emission from the 
dust and debris disks around nearby stars 
(Heinrichsen \etal 1999; Schneider 2001)
must be interpreted in the 
context of the formation and evolution of such disks. 
They are very likely 
to be the outcome of the same collisional fragmentation and 
accumulation of massive 
rocky or icy bodies that form terrestrial planets.

It is widely believed that the formation of Earth-type planets 
proceeded via agglomeration of large number of planetesimals --- 
asteroid or comet-like
rocky or icy bodies. The theory of this process
was pioneered by Safronov (1972) who was the first to point out the 
importance of the evolution of the dynamical properties of the 
planetesimal disk for the 
evolution of its mass distribution. The discovery of a rapid ``runaway'' 
mode of planetary accretion  by Wetherill \& Stewart 
(1989) has made this issue even more important since this mechanism 
relies on the redistribution of the energy of planetesimal epicyclic 
motion between populations with different masses 
(Wetherill \& Stewart 1993; Kenyon \& Luu 1998; Inaba \etal 2001).
Significant progress has been made recently in understanding the dynamical 
evolution of {\it homogeneous} planetesimal disks --- 
disks where gradients 
of surface density or dynamical properties (such as the dispersions of 
eccentricity and inclination of various planetesimal populations) are 
absent (Ida 1990; Stewart \& Ida 2000, hereafter SI; Ohtsuki \etal 2002). 
This assumption should be very good during the initial stages of planetesimal
growth when there are no massive bodies in the disk. 
However as coagulation proceeds and 
planetary embryos --- precursors of terrestrial planets --- emerge
this assumption runs into problems. It was first demonstrated by 
Ida \& Makino (1993) using N-body simulations  and then confirmed 
using orbit integrations (Tanaka \& Ida 1997, 1999) 
that under a variety of conditions 
massive protoplanetary embryos would tend
to repel planetesimal orbits, clearing out an annular gap in the 
disk around them.  This process introduces a new spatial dimension
into the problem and makes it much harder to treat.

N-body simulations are not good tools to
study the details of this process. The primary reason is that they become
too time-consuming when one needs to follow the spatial and dynamical 
properties of a many-body system for many orbital periods. Moreover, the number
of planetesimals which they can handle is not very large (less than $10^4$)
which precludes the consideration of realistic planetesimal disks containing
huge number of bodies with masses spanning an enormous range ---
from one meter rocks to $100$ km planetesimals. Finally, one would need to
perform a large number of such simulations to explore the whole 
space of physical parameters relevant for protoplanetary disk evolution.
Similar problems, although less severe, plague the
performance of methods based of the direct integration of planetesimal 
orbits using some simplifying assumptions (Tanaka \& Ida 1997, 1999).

The large number of bodies under consideration poses 
no problem for the methods of kinetic theory. In this approach the
disk is split into a set of planetesimal populations and each of them
is characterized by three functions of position: the surface density and 
dispersions of eccentricity and inclination. Application of this method
for studying planetesimal coagulation and the evolution of dynamical 
properties in homogeneous disks has proven to be very useful
(Wetherill \& Stewart 1989; Kenyon \& Luu 1998). 
It can also be easily extended to the case of inhomogeneous disks by allowing 
the planetesimal properties to vary within the disk. This approach 
allows one to study many physical mechanisms important for the 
coagulation process --- 
gravitational interactions between planetesimals increasing their 
random motion, gas drag and inelastic collisions damping them, migration, 
fragmentation, etc. The use of this statistical approach for exploring 
inhomogeneous disks was first undertaken by Petit \& H\'enon 
(1987a, 1987b, 1988) in their studies of planetary rings. Physical conditions 
in planetary rings (high optical depth and frequent inelastic collisions)
are very different from those which are thought to exist in protoplanetary 
disks. Thus we cannot directly apply the methods of Petit \& H\'enon
to study planetesimal disks, 
but the spirit in which their investigation was carried out
can be largely preserved.

In this paper we consider  the evolution of 
{\it inhomogeneous} planetesimal disks
caused by interactions between planetesimals using 
conventional methods of statistical mechanics (Lifshitz \& Pitaevskii 1981).
The effects of the planetary embryos
on the disk evolution will be investigated later in Rafikov (2002; hereafter 
Paper II).
We will concentrate on one of the most important processes
going on in these disks --- mutual gravitational scattering of 
planetesimals --- although our rather
general approach allows one to treat other relevant phenomena as well.
This process  can proceed in two distinct regimes 
depending on the amplitude of planetesimal random motion. The
gravitational 
attraction between two planetesimals with masses $m_1$ and $m_2$
becomes stronger than the tidal field 
of the central star of mass $M_c$ when their mutual 
separation is less than their 
Hill (or tidal, or Roche) radius $r_H$, defined as
\begin{eqnarray}
r_H=a_0\left(\frac{m_1+m_2}{M_c}\right)^{1/3},
\label{eq:Hill_radius}
\end{eqnarray}
where $a_0$ is the distance from the central star.
When the random velocities of the epicyclic
motion of interacting planetesimals are smaller than $\sim \Omega r_H$
($\Omega=\sqrt{GM_c/a_0^3}$ is the disk orbital frequency at $a_0$) their 
relative approach velocities are 
small and close interactions can lead to a considerable change of the orbital 
elements of planetesimals. This velocity regime is called shear-dominated
(or ``cold'').
It should be contrasted with the other extreme --- the so called
dispersion-dominated  (``hot'') regime which occurs when planetesimal velocity 
dispersions are bigger than $\sim \Omega r_H$. In this latter case 
scattering is typically weak which often allows analytical
treatment of this velocity regime.

The development of planetesimal disk inhomogeneities driven by a 
protoplanetary embryo was explored in Rafikov (2001; hereafter Paper I) 
assuming that shear-dominated scattering
of planetesimals prevails. It was also assumed in this study that the 
dynamical properties of planetesimals do not evolve as a result of 
scattering and that the disk always stays dynamically cold.  
Planetesimal-planetesimal interactions 
play the role of effective viscosity in the disk, 
and tend to homogenize it and close up any gap.
Nevertheless, this study 
demonstrated that gap formation is the
 natural outcome of the embryo-planetesimal interaction
when the embryo is massive enough. These interactions were local in 
character because in the shear-dominated case 
only planetesimals on orbits separated by no more 
than several $r_H$ (corresponding to their encounter)
were able to approach each other closely. 

It is however more likely that the 
planetesimal-planetesimal gravitational scattering in realistic 
protoplanetary
disks proceeds in the dispersion-dominated (rather than 
shear-dominated) regime, at least on the late stages of disk evolution
(see \S \ref{sect:FP_expansion}). 
In this case the
evolution of planetesimal random motion can strongly
affect the growth rate of protoplanetary embryos. It is also tightly 
coupled to the evolution of spatial distribution of planetesimals because
any change of the energy of epicyclic motion comes at the expense of 
the orbital energy of planetesimals.

The treatment of the dispersion-dominated 
case is complicated by the fact that planetesimals 
in this regime can explore 
different regions of the disk in the course of their epicyclic
motion. This makes disk evolution a nonlocal process. On the other hand, as we 
have said before there are natural simplifications 
which are valid in the dispersion-dominated regime.
These include the two-body scattering approximation (relative
velocities are high), Fokker-Planck type expansions
(scattering is weak), and so on.

Our paper is organized as follows. In \S \ref{sect:master_equation}
we derive general equations for the planetesimal surface density 
and velocity dispersion evolution. We do this using the Hill
approximation which is briefly described in \S \ref{subsect:Hill_eq}. A
Fokker-Planck expansion of the evolution equations,
valid in the dispersion-dominated regime, is performed
in \S \ref{sect:FP_expansion}.
In \S \ref{sect:scattering_coefficients} we derive analytical expressions for 
the scattering coefficients used in these equations. We conclude
with a brief summary of our results in \S \ref{sect:Summary}. 
Some technical details of the calculations and derivations can be 
found in appendices.


\section{Equations of evolution of planetesimal disk properties.}
\label{sect:master_equation}


Throughout the paper we will assume disk to be axisymmetric.
All the following calculations also assume a disk with a Keplerian 
rotation law, although they could be easily extended to the case of 
an arbitrary rotation law. 

As we have mentioned before planetesimal disks must have contained a huge
number of constituent bodies with a wide range of masses. 
Thus we use a kinetic 
approach and characterize the state of the disk at any 
moment of time $t$ by
a set of three averaged parameters for each mass population: the surface
number density of planetesimals 
and their velocity dispersions in 
the horizontal and vertical directions  
at every point ${\bf r}$ in the disk.   


When the orbit's
eccentricity and inclination are small, it is convenient to work in the
guiding center approximation, when the elliptic motion is represented as
an epicyclic motion of the body on a small ellipse in horizontal plane
with the radial dimension $a e$ (and oscillatory motion in the 
vertical direction with the amplitude $a i$);
the  center of this ellipse
performs circular motion with a semimajor axis $a$.
In the case of Keplerian rotation law frequencies of all these motions
are the same and equal to the angular frequency of Keplerian motion.
In this approximation the roles of longitude of ascending
node and argument of pericentre are played by two constant phase
angles $\omega$ and $\tau$ which characterize the 
position of body on its epicycle at a given moment of time.
Random motion is defined as the motion relative to the circular orbit
of the guiding center, i.e. is represented by the epicyclic motion.
Eccentricity and inclination thus measure the magnitude of 
the random motion and we will take them to represent planetesimal's 
kinematic properties.

The spatial distribution of planetesimals can be characterized by 
two kinds of surface density. One is the {\it surface 
density of guiding centers} of planetesimals
$N(a)$ and is defined such that $N(a)2\pi a da$ is the number
of planetesimals with guiding centers between $a$ and $a+da$. 
It is a function of planetesimal semimajor axes only. 

Another parametrization is the 
{\it instantaneous surface density} $N^{inst}(r)$,
 defined such that $N^{inst}(r)2\pi r dr$ is the number of
planetesimals within the disk surface element $2\pi r dr$ (we will denote 
instantaneous distance from the central star as $r$ to distinguish it
from $a$). The relation between these two surface densities 
can only be computed if the distribution of 
planetesimal eccentricities is known
(we calculate an appropriate conversion between $N$ and $N^{inst}$ in 
Appendix \ref{app:n_n_conversion}). These surface 
densities coincide only when planetesimals are on circular orbits 
(which was the case in Paper I). 
We will mostly work with the surface 
density of the guiding centers and not the instantaneous surface density.


\subsection{Hill equations.}
\label{subsect:Hill_eq}


In the epicyclic approximation, which is valid when 
$e\ll 1$ and $i\ll 1$, it is convenient to
 introduce the Cartesian coordinate system $x,y,z$ with axes in 
the $r,\varphi,z$ directions of the corresponding cylindrical system
and origin at some reference distance $a_0$ from the central star.
This system rotates around the central star with 
angular velocity $\Omega_0=\Omega(a_0)$.
In these coordinates the unperturbed motion of the body is given by 
(Goldreich \& Lynden-Bell 1965; H\'enon \& Petit 1986)
\begin{eqnarray}
&& x=h a_0-e a_0\cos(\Omega_0 t+\tau)+O(e^2,i^2),\nonumber\\
&& y=-\frac{3}{2}h a_0 (\Omega_0 t-\lambda)
+2e a_0\sin(\Omega_0 t+\tau)+O(e^2,i^2),\nonumber\\
&& z=i a_0\sin(\Omega_0 t+\omega)+O(e^2,i^2),
\label{eq:unpert_motion}
\end{eqnarray} 
where $\tau$ and $\omega$ are the constant phase angles 
we have mentioned before
and $\lambda$ is the origin of time (it is usually unimportant and 
can be set equal to $0$). Dimensionless relative semimajor axis 
$a$ of
the body (the location of its guiding center) $h$ is defined as 
\begin{eqnarray}
h=\frac{a-a_0}{a_0}
\label{eq:h_def}
\end{eqnarray} 
Instead of the pairs $e,\tau$ and 
$i,\omega$ it is sometimes convenient to use the eccentricity 
and inclination {\it vectors} ${\bf e}$ and ${\bf i}$:
\begin{eqnarray}
&& {\bf e}=(e_x,e_y)=
(e\cos\tau,e\sin\tau),\nonumber\\
&& {\bf i}=(i_x,i_y)=
(i\cos\omega,i\sin\omega).
\label{eq:reduced_e_i}
\end{eqnarray}

We can also use other natural simplifications.
One of them is motivated by the
assumption that the masses of planetesimals $m_j$
are much smaller than the mass of the central star
$M_c$ --- a condition which is always fulfilled in planetesimal disks.
In this case it was demonstrated by H\'enon \& Petit (1986) that the 
equations describing the $3$-body interaction (star and 
planetesimals 1 and 2) can be significantly simplified.
In particular they have demonstrated that relative motion
of interacting bodies (${\bf r}={\bf r}_1-{\bf r}_2$)
and the motion of their barycenter 
[${\bf r}_b=(m_1{\bf r}_1+m_2{\bf r}_2)/(m_1+m_2)$]
separate from each other. In the course of $3$-body interactions
the barycenter motion is not affected at all and remains in unperturbed 
epicyclic motion at all times. 

We define the relative 
Hill coordinates 
of the two interacting bodies $(\tilde x,\tilde y,\tilde 
z)$ with masses $m_1\ll M_c$ and $m_2\ll M_c$ as
\begin{eqnarray}
\tilde x=(x_1-x_2)/r_H,~~~
\tilde y=(y_1-y_2)/r_H,~~~
\tilde z=(z_1-z_2)/r_H.
\label{eq:Hill_eq}
\end{eqnarray}
Here  $r_H$ is the Hill radius defined by equation (\ref{eq:Hill_radius}).
Instead of $m$ we will often use $\mu=m/M_c$ [with this notation 
the Hill radius can be written as $r_H=a_0(\mu_1+\mu_2)^{1/3}$].
Note that the Hill radius is only meaningful for a {\it pair} of bodies and
its definition is similar to the definition of tidal radius. 
We will also define {\it reduced} eccentricity and inclination vectors 
$\tilde {\bf e}=a_0({\bf e}_1-{\bf e}_2)/r_H, \tilde {\bf i}=
a_0({\bf i}_1-{\bf i}_2)/r_H$ and the reduced impact
parameter $\tilde h=a_0(h_1-h_2)/r_H$. 

 With these 
definitions the equations of relative motion can be written as
(H\'enon \& Petit 1986; Hasegawa \& Nakazawa 1990)
\begin{eqnarray}
&& \frac{d^2\tilde x}{d T^2}-2\frac{d\tilde y}{d T}-3\tilde x=-
\frac{\partial \phi_p}{\partial \tilde x},\label{eq:xmot}\\
&& \frac{d^2\tilde y}{d T^2}+2\frac{d\tilde x}{d T}=
-\frac{\partial \phi_p}{\partial \tilde y},\label{eq:ymot}\\
&& \frac{d^2\tilde z}{d T^2}+ z=
-\frac{\partial \phi_p}{\partial \tilde z}\label{eq:zmot},
\end{eqnarray}
where $\phi_p$ is a (dimensionless) 
potential of interaction between the two bodies
and $T=\Omega_0 t$. For the Newtonian gravitational potential  between
planetesimals
\begin{eqnarray}
\phi_p({\bf r})=-(\tilde x^2+\tilde y^2+\tilde z^2)^{-1/2}.
\label{eq:potential}
\end{eqnarray}
An important feature of these equations is that they do not contain 
any parameters of the interacting bodies (such as their masses, etc.). 
All the physically important
information is embedded into the definitions of Hill radius and relative
coordinates. Also, the outcome of the interaction between the two bodies 
depends only on their {\it relative} orbital parameters.
This universality is a very useful property which we will 
widely exploit later on. 
Equations (\ref{eq:xmot})-(\ref{eq:zmot}) possess an integral of motion ---
the Jacobi constant
\begin{eqnarray}
J=\left(\frac{d\tilde x}{dT}\right)^2+
\left(\frac{d\tilde y}{dT}\right)^2+
\left(\frac{d\tilde z}{dT}\right)^2
-3\tilde x^2+\tilde z^2+2\phi_p.
\label{eq:Jacobi_xyz}
\end{eqnarray}
Conservation of this quantity poses an 
important constraint on the changes of the orbital elements of 
interacting bodies in the course of their gravitational interaction.

One can easily see that when the interaction is absent equations 
(\ref{eq:unpert_motion}) represent a solution of the system 
(\ref{eq:xmot})-(\ref{eq:zmot}) if we replace all quantities 
in (\ref{eq:unpert_motion}) by
their relative (and scaled by $r_H$)
analogs. But in  the presence of the interaction, 
orbital parameters
are no longer constants and will evolve with time. However, we can still 
represent epicyclic motion by equations (\ref{eq:unpert_motion}) keeping
in mind that $\tilde {\bf e}, \tilde {\bf i}, \tilde h$, etc. are 
now {\it instantaneous} osculating values of the 
orbital elements. Their evolution is governed by the following set of 
equations (Goldreich \& Tremaine 1980; Hasegawa \& Nakazawa 1990):
\begin{eqnarray}
&& \frac{d\tilde h}{d T}=-2\frac{\partial \phi_p}{\partial \tilde y},~~~
\frac{d\lambda}{d T}=\frac{4}{3\tilde h}
\frac{\partial \phi_p}{\partial \tilde x}
-\frac{2}{\tilde h}(T-\lambda)\frac{\partial \phi_p}{\partial \tilde y},
\nonumber\\
&& \frac{d\tilde e_x}{d T}=-\sin T\frac{\partial \phi_p}{\partial \tilde x}-
2\cos T\frac{\partial \phi_p}{\partial \tilde y},~~~
\frac{d\tilde e_y}{d T}=-\cos T\frac{\partial \phi_p}{\partial \tilde x}+
2\sin T\frac{\partial \phi_p}{\partial \tilde y},\nonumber\\
&& \frac{d\tilde i_x}{d T}=
-\cos T \frac{\partial \phi_p}{\partial \tilde z},~~~
\frac{d\tilde i_y}{d T}=\sin T\frac{\partial \phi_p}{\partial \tilde z}.
\label{eq:osculating}
\end{eqnarray}

The relative velocity $\dot{\bf r}$ of planetesimal motion in Hill coordinates 
(dimensional)
is given  by
[see (\ref{eq:unpert_motion})]
\begin{eqnarray}
\dot x= \tilde e r_H\sin(T+\tau),~~~
\dot y=-\frac{3}{2}\tilde h r_H +2\tilde e r_H\cos(T+\tau),~~~
\dot z= \tilde i r_H\cos(T+\omega).
\label{eq:Hill_vels}
\end{eqnarray}
In many applications it is better to use the velocity  defined 
{\it relative to the local velocity of the circular motion}. Clearly 
this velocity ${\bf v}$ is related to ${\bf \dot r}$ as
\begin{eqnarray}
{\bf v}={\bf \dot r}+\frac{3}{2}x{\bf n}_y,
\label{eq:Hill_to_local}
\end{eqnarray}
where ${\bf n}_y$ is a unit vector in the $y$-direction. This definition
implies that
\begin{eqnarray}
v_x=\dot x=\tilde e r_H\sin(T+\tau),~~~
v_y=\dot y+\frac{3}{2}x=\frac{\tilde e r_H}{2}\cos(T+\tau),~~~
v_z=\dot z= \tilde i r_H\cos(T+\omega).
\label{eq:local_vels} 
\end{eqnarray}
With the use of equations (\ref{eq:unpert_motion}) and
(\ref{eq:local_vels}) one can express $e$, $i$ and $h$ as functions
of ${\bf r}$ and ${\bf v}$ which will later be used in 
\S \ref{sect:scattering_coefficients}. 
Using (\ref{eq:unpert_motion}) and (\ref{eq:Hill_vels}) we can rewrite 
the expression (\ref{eq:Jacobi_xyz}) for the Jacobi constant in the 
following form:
\begin{eqnarray}
J=\tilde e^2+\tilde i^2-\frac{3}{4}\tilde h^2+2\phi_p
\label{eq:Jacobi}
\end{eqnarray}
(our definition of the Jacobi constant may differ from others by a
constant additive and/or multiplicative factors, cf. 
Hasegawa \& Nakazawa 1990).


\subsection{Derivation of the master equation for the distribution function.}
\label{subsect:dist_func}


Based on the previous discussion it is natural to describe 
the position and velocity state of a planetesimal with mass $m_j$
by its guiding center radius $h_j$ (relative to the reference radius $a_0$), 
vector eccentricity ${\bf e}_j$ and inclination  ${\bf i}_j$\footnote{
These variables are dimensionless but not scaled by $r_H$ as in \S 
\ref{subsect:Hill_eq}.} 
(note that this is not a canonical set of 
variables!). 
The five-dimensional vector ${\bf \Gamma}_j=(h_j,{\bf e}_j,{\bf i}_j)$
will be used to concisely denote this state and 
$d{\bf \Gamma}_j=dh_j d{\bf e}_j d{\bf i}_j$ will denote a phase space volume 
element around ${\bf \Gamma}_j$.

We set $N(m,h,t) dm$ to be the dimensionless surface number density 
of planetesimal guiding centers in the mass interval $(m,m+dm)$.
It differs from its dimensional analog $N(a)$ by a factor of $a_0^2$. 
We introduce the notion of a planetesimal distribution function
(PlDF) $f(m,{\bf \Gamma},t)$ such that $2\pi 
f(m,{\bf \Gamma},t)d{\bf \Gamma}dm$ is the number
of planetesimals in phase space volume $d{\bf \Gamma}$
and in mass interval $(m,m+dm)$. Clearly,
\begin{equation}
\int\int f(m,{\bf \Gamma},t)d{\bf e}d{\bf i}=N(m,h,t),
\label{eq:normalization}
\end{equation}
where the integration is performed over the 4-dimensional space of vector
eccentricity. For the purpose of brevity we will further 
use $f_j({\bf \Gamma},t)$ and $N_j(h,t)$ instead of 
$f(m_j,{\bf \Gamma}_j,t)$ and  $N(m_j,h,t)$.

In what follows we will consider the distribution function of planetesimal
vector eccentricities and inclinations
to be Gaussian (or Rayleigh, or Schwarzschild). 
In other words, we assume that
$f_j({\bf \Gamma})d{\bf \Gamma}=N(h)\psi_j({\bf e},{\bf i},h)d{\bf \Gamma}$, 
where
\begin{equation}
\psi_j({\bf e},{\bf i},h)=\frac{1}{4\pi^2\sigma_{e j}^2(h)
\sigma_{i j}^2(h)}\exp\left[-\frac{{\bf e}^2}{2\sigma_{e j}^2(h)}
-\frac{{\bf i}^2}{2\sigma_{i j}^2(h)}\right].
\label{eq:Gauss_DF}
\end{equation}
Here $\sigma_{e j},
\sigma_{i j}$ are the r.m.s. eccentricities and inclinations 
of planetesimals of mass $m_j$ at point $h$, given by
\begin{eqnarray}
\sigma_{e j}=\langle{\bf e}_j^2\rangle^{1/2}/\sqrt{2},~~~ 
\sigma_{i j}=\langle{\bf i}_j^2
\rangle^{1/2}/\sqrt{2}.
\end{eqnarray}
Greenzweig \& Lissauer (1992) and Ida \& Makino (1992) 
have demonstrated that this assumption 
of Gaussian distribution is very good 
in their N-body simulations for large values of planetesimal
eccentricities and inclinations [see also Ohtsuki (1999) 
for the discussion of
the validity of distribution (\ref{eq:Gauss_DF}) in circumplanetary disks].

Consider now planetesimals with masses $m_1$ (type 1)
and $m_2$ (type 2) (we usually denote the particle under 
consideration by subscript 1  and scattered particles by subscript 2). 
Scattering 
can bring planetesimals of type 1 into the 
phase volume element $d{\bf \Gamma}$
and can also scatter particles out of it. The number of encounters 
in time $dt$ between 
particles of type 1 {\it initially} in the phase volume 
element $d{\bf \Gamma}_a$
and those of type 2 initially in $d{\bf \Gamma}_b$ is
\begin{eqnarray}
dN_{coll}=\nu({\bf \Gamma}_a,{\bf \Gamma}_b)
f_1({\bf \Gamma}_a)f_2({\bf \Gamma}_b)
d{\bf \Gamma}_a d{\bf \Gamma}_b dt,
\label{eq:Ncoll}
\end{eqnarray}
where $\nu({\bf \Gamma}_a,{\bf \Gamma}_b)$ is the rate of encounters.

Let us introduce the differential probability 
$P({\bf \Gamma}_a,{\bf \Gamma}_b,\Delta{\bf \Gamma}_a)$ such that 
$P({\bf \Gamma}_a,{\bf \Gamma}_b,\Delta{\bf \Gamma}_a)
d\left(\Delta{\bf \Gamma}_a\right)$ is the probability that particle 1  
changes its state from ${\bf \Gamma}_a$ to the region $d\Delta{\bf \Gamma}_a$ 
around 
${\bf \Gamma}_a+\Delta{\bf \Gamma}_a$
in an encounter with particle of type 2,
if before the collision they were in states 
${\bf \Gamma}_a, {\bf \Gamma}_b$ respectively.
We can then write down the evolution 
equation for PlDF of particles of type 1:
\begin{eqnarray}
&& \frac{\partial f_1({\bf \Gamma})}{\partial t}+
{\bf \dot\Gamma}\frac{\partial f_1({\bf \Gamma})}{\partial {\bf \Gamma}}=
\int\limits_0^\infty dm_2\int d{\bf \Gamma}_a d{\bf \Gamma}_b
\nonumber\\
&& \times\bigg[\nu({\bf \Gamma}_a,{\bf \Gamma}_b)
f_1({\bf \Gamma}_a)f_2({\bf \Gamma}_b)
P({\bf \Gamma}_a,{\bf \Gamma}_b,{\bf \Gamma}-{\bf \Gamma}_a)-
\nu({\bf \Gamma},{\bf \Gamma}_b)f_1({\bf \Gamma})f_2({\bf \Gamma}_b)
P({\bf \Gamma},{\bf \Gamma}_b,{\bf \Gamma}_a-{\bf \Gamma})\bigg].
\label{eq:general_nonlocal_master}
\end{eqnarray}
Here the second term in the l.h.s. represents evolution
caused by reasons other than mutual gravitational scattering (e.g. gas drag, 
migration, non-Keplerian gravitational forces, etc.). The r.h.s. of 
(\ref{eq:general_nonlocal_master}) is the collision integral 
for scattering between planetesimals; its first term represents
particles entering $d{\bf \Gamma}$, while the second one represents 
planetesimals leaving $d{\bf \Gamma}$.
This is the most general form of evolution equation, which will be further 
simplified using additional assumptions.


\subsection{Local approximation.}
\label{subsect:local}


When the masses of the interacting bodies are much smaller than $M_c$
their mutual Hill radius is much smaller than 
the distance to the central star $a_0$. We will also assume that the 
sizes of epicycles of individual bodies are small compared to $a_0$. 
As a result, we can safely use the Hill formalism (\S \ref{subsect:Hill_eq}) 
to describe planetesimal scattering.

The local character of the interaction 
means that  the encounter
rate $\nu({\bf \Gamma}_a,{\bf \Gamma}_b)$ is determined by the local 
shear and can be expressed as 
\begin{eqnarray}
\nu({\bf \Gamma}_a,{\bf \Gamma}_b)=
|\Omega(h_a)-\Omega(h_b)|=2 |A| |h_a-h_b|,
\end{eqnarray}
where $A=(r/2)(d\Omega/dr)$ is a function characterizing shear in the disk
(Oort's $A$ constant), 
$A=-(3/4)\Omega_0$ for a Keplerian rotation law.
Using this expansion we can rewrite equation
(\ref{eq:general_nonlocal_master}) as
\begin{eqnarray}
&& \frac{\partial f_1({\bf \Gamma})}{\partial t}+
{\bf \dot\Gamma}\frac{\partial f_1({\bf \Gamma})}{\partial {\bf \Gamma}}=
2|A|\int\limits_0^\infty dm_2\int d{\bf \Gamma}_a d{\bf \Gamma}_b
\nonumber\\
&& \times\bigg[|h_a-h_b|f_1({\bf \Gamma}_a)f_2({\bf \Gamma}_b)
P({\bf \Gamma}_a,{\bf \Gamma}_b,{\bf \Gamma}-{\bf \Gamma}_a)-
|h-h_b|f_1({\bf \Gamma})f_2({\bf \Gamma}_b)
P({\bf \Gamma},{\bf \Gamma}_b,{\bf \Gamma}_a-{\bf \Gamma})\bigg].
\label{eq:general_local_master}
\end{eqnarray}

We can easily integrate the second term 
over $d{\bf \Gamma}_a$ because  it is clear that
\begin{equation}
\int d{\bf \Gamma}_a P({\bf \Gamma},{\bf \Gamma}_b,
{\bf \Gamma}_a-{\bf \Gamma})=1, 
\end{equation} 
since this integral represents the probability of scattering planetesimal 1
(initially at ${\bf \Gamma}$)
{\it anywhere} in the course of interaction with planetesimal 2 (initially at 
${\bf \Gamma}_b$). Then we can also integrate $f_2({\bf \Gamma}_2)$ over
$d{\bf e}_2d{\bf i}_2$ to find
\begin{equation}
\int d{\bf \Gamma}_a d{\bf \Gamma}_b
|h-h_b|f_1({\bf \Gamma})f_2({\bf \Gamma}_b)
P({\bf \Gamma},{\bf \Gamma}_b,{\bf \Gamma}_a-{\bf \Gamma})=
\int_{-\infty}^\infty dh_b
|h-h_b|f_1({\bf \Gamma})N_2(h_b).
\label{eq:sec_term}
\end{equation}

To deal with the first term in the r.h.s. of (\ref{eq:general_local_master}) 
we first recall that in the Hill approximation all 
scattering properties depend only on {\it relative} coordinates 
and velocities of planetesimals. This means that
\begin{equation}
P({\bf \Gamma}_a,{\bf \Gamma}_b,\Delta{\bf \Gamma}_a)=
P({\bf \Gamma}_a-{\bf \Gamma}_b,\Delta{\bf \Gamma}_a),
\label{eq:property}
\end{equation}
Now, let us introduce the 
new relative variable
\begin{eqnarray} 
{\bf \tilde\Gamma}_r=
(\tilde h,{\bf \tilde e}_r,{\bf \tilde i}_r)
=\frac{{\bf \Gamma}_a-{\bf \Gamma}_b}{(\mu_1+\mu_2)^{1/3}}=
\frac{1}{(\mu_1+\mu_2)^{1/3}}
\left(h_a-h_b,{\bf e}_a-{\bf e}_b,{\bf i}_a-{\bf i}_b\right),
\label{eq:relative_var}
\end{eqnarray}
where ${\bf \Gamma}_a$ and  ${\bf \Gamma}_a$ characterize the orbital
elements of particles $1$ and $2$ correspondingly.
The change of ${\bf \tilde\Gamma}_r$ in the 
course of an encounter is $\Delta{\bf \tilde\Gamma}_r$, and depends only
on ${\bf \Gamma}_a$ and ${\bf \Gamma}_b$ through the combination
${\bf \tilde\Gamma}_r$. Then, bearing in mind conservation 
of the center of mass coordinate $m_1{\bf \Gamma}_a+m_2{\bf \Gamma}_b$,
we obtain that ${\bf \Gamma}_a, {\bf \Gamma}_b$ can be expressed in terms of 
${\bf \Gamma}$ [in the first term of (\ref{eq:general_local_master}) 
this is the value of ${\bf \Gamma}_a$ after the encounter], 
${\bf \tilde\Gamma}_r$,
and $\Delta{\bf \tilde\Gamma}_r$ as
\begin{eqnarray}
{\bf \Gamma}_a={\bf \Gamma}-\frac{\mu_2}{\mu_1+\mu_2}(\mu_1+\mu_2)^{1/3}
\Delta{\bf \tilde\Gamma}_r,~~~
{\bf \Gamma}_b={\bf \Gamma}-(\mu_1+\mu_2)^{1/3}
\left(\frac{\mu_2}{\mu_1+\mu_2}\Delta{\bf \tilde\Gamma}_r+
{\bf \tilde\Gamma}_r\right).
\label{eq:var_change}
\end{eqnarray}

Instead of the probability $P({\bf \Gamma}_a,{\bf \Gamma}_b,
{\bf \Gamma}-{\bf \Gamma}_a)d\left(\Delta{\bf \Gamma}_a\right)$ 
we will work with the probability
of scattering in relative coordinates 
$\tilde P_r({\bf \tilde\Gamma}_r,\Delta{\bf \tilde\Gamma}_r)
d(\Delta{\bf \tilde\Gamma}_r)$. 
From the equation  (\ref{eq:var_change}) one can easily find that
\begin{eqnarray}
\tilde P_r({\bf \tilde\Gamma}_r,\Delta{\bf \tilde\Gamma}_r)=
\frac{\mu_2^5}{(\mu_1+\mu_2)^{10/3}}
P({\bf \Gamma}_a-{\bf \Gamma}_b,\Delta{\bf \Gamma}_a).
\end{eqnarray}  
This new probability function is independent of the masses of the 
interacting 
particles. It also follows from the properties of the Hill equations that
\begin{equation}
\tilde P_r({\bf \tilde\Gamma}_r,\Delta{\bf \tilde\Gamma}_r)=
\delta\left[\Delta{\bf \tilde\Gamma}_r-
\Delta{\bf \tilde\Gamma}_{sc}({\bf \tilde\Gamma}_r)\right],
\label{eq:delta_fun}
\end{equation}
i.e. the initial state of the scattering planetesimals 
unambiguously determines their final state through the scattering function
$\Delta{\bf \tilde\Gamma}_{sc}({\bf \tilde\Gamma}_r)$.

Using these definitions 
we can finally put the
PlDF evolution equation 
(\ref{eq:general_local_master}) in the following form:
\begin{eqnarray}
&& \frac{\partial f_1({\bf \Gamma})}{\partial t}+
\frac{d {\bf \Gamma}}{d t}
\frac{\partial f_1({\bf \Gamma})}
{\partial {\bf \Gamma}}=-
2|A|\int\limits_0^\infty dm_2(\mu_1+\mu_2)^{2/3}
\int\limits_{-\infty}^\infty d\tilde h|\tilde h|
\nonumber\\
&& \times\bigg[f_1({\bf \Gamma})
N_2(h-(\mu_1+\mu_2)^{1/3}\tilde h)
\label{eq:general_hill_master}\\
&& -(\mu_1+\mu_2)^{4/3}
\int d{\bf \tilde e}_r d{\bf \tilde i}_r d(\Delta{\bf \tilde\Gamma}_r)
f_1({\bf \Gamma}_a)f_2({\bf \Gamma}_b)
P_r({\bf \tilde\Gamma}_r,\Delta{\bf \tilde\Gamma}_r)\bigg],
\nonumber
\end{eqnarray}
with ${\bf\Gamma},{\bf\Gamma}_a,{\bf \Gamma}_b,{\bf \tilde\Gamma}_r$ and
$\Delta{\bf \tilde\Gamma}_r$
related by equations (\ref{eq:relative_var}) and (\ref{eq:var_change}).

We will use this general equation for the evolution of the PlDF
to obtain formulae describing the behavior of the spatial 
and kinematic properties of planetesimal disk.


\subsection{Evolution of surface density.}
\label{subsect:evol_surf}


In this paper we 
will be mostly interested in the evolution of the planetesimal 
disk due to the gravitational scattering between planetesimals. 
Thus, we set the second
term in the l.h.s. of equation (\ref{eq:general_hill_master}) 
to $0$ (it can be easily reinstated when  the need arises). 
At this stage we also replace subscripts ``a'' and ``b'' with ``1'' and
``2'' since now ${\bf \Gamma}_a$ and ${\bf \Gamma}_b$ have the only 
meaning of initial orbital elements of particles 1 and 2.

To obtain the equation of evolution of $N_1(h,t)$ let us multiply equation 
(\ref{eq:general_hill_master}) by $d{\bf e} d{\bf i}$ and integrate. 
Using equation 
(\ref{eq:normalization}) we can easily do this with the l.h.s. and the 
first term of 
the r.h.s. To treat the second term in the r.h.s. of
(\ref{eq:general_hill_master}) 
we substitute the distribution function (\ref{eq:Gauss_DF}) into
equation (\ref{eq:general_hill_master}) and integrate over $d{\bf e}$ using 
(\ref{eq:var_change}) and the following identity:
\begin{eqnarray}
\int \frac{d{\bf e}}{4\pi^2\sigma_{e 1}^2
\sigma_{e 2}^2}\exp\left[-\frac{{\bf e}_1^2}
{2\sigma_{e 1}^2}
-\frac{{\bf e}_2^2}{2\sigma_{e 2}^2}\right]
=\frac{1}{2\pi\left(\sigma_{e 1}^2+\sigma_{e 2}^2\right)}
\exp\left[-\frac{{\bf e}_r^2}
{2\left(\sigma_{e 1}^2+\sigma_{e 2}^2\right)}\right]
\equiv \psi_r\left({\bf e}_r\right),
\label{eq:f_rel_def}
\end{eqnarray}
with $\psi_r\left({\bf \tilde e}_r\right)$ being the distribution function
of {\it relative} eccentricity as defined by (\ref{eq:f_rel_def}).
An analogous identity defines the distribution function
of relative inclination $\psi_r({\bf \tilde i}_r)$.
These identities show that the distribution function for
the {\it relative} motion of planetesimals with 
r.m.s. eccentricities and inclinations $\sigma_{e 1},
\sigma_{i 1}$ and $\sigma_{e 2},\sigma_{i 2}$ 
is also Gaussian, with dispersions of the relative eccentricities and 
inclinations given by 
\begin{equation}
\sigma_{e r}^2=\sigma_{e 1}^2+\sigma_{e 2}^2,~~~~~
\sigma_{i r}^2=\sigma_{i 1}^2+\sigma_{i 2}^2.
\label{eq:rel_ecc}
\end{equation}

Using this  result we are able to write the following 
equation for the evolution of
$N_1(h)$:
\begin{eqnarray}
&& \frac{\partial N_1(h)}{\partial t}=-2|A|\int\limits_0^{\infty}dm_2
(\mu_1+\mu_2)^{2/3}
\int\limits_{-\infty}^{\infty}d\tilde h |\tilde h|
\Bigg\{N_1(h)N_2[h-(\mu_1+\mu_2)^{1/3} \tilde h]\nonumber \\
&& -\left.
(\mu_1+\mu_2)^{4/3}
\int d{\bf \tilde e}_r d{\bf \tilde i}_r d(\Delta{\bf \tilde\Gamma}_r)
\psi_r({\bf e}_r,\sigma_{e r})\psi_r({\bf i}_r,\sigma_{i r})N_1(h_1)N_2(h_2)
P_r({\bf \tilde\Gamma}_r,\Delta{\bf \tilde\Gamma}_r)\right\},
\label{eq:interm_n_eq}
\end{eqnarray}
 where
\begin{eqnarray}
&& \sigma_{e r}^2(h_1,h_2)=\sigma_{e 1}^2(h_1)+
\sigma_{e 2}^2(h_2),~~~
\sigma_{i r}^2(h_1,h_2)=\sigma_{i 1}^2(h_1)+ 
\sigma_{i 2}^2(h_2),\\
&& h_1=h+D(\Delta \tilde h),~~~~~~~
h_2=h+D(\Delta \tilde h)-(\mu_1+\mu_2)^{1/3} \tilde h,
\label{r_s}\nonumber \\  
&& \mbox{with}~~~~~~
D(\Delta \tilde h)=-\frac{\mu_2 }{(\mu_1+\mu_2)^{2/3}}\Delta \tilde h.
\label{eq:defs_Dh}
\end{eqnarray}

In the last term of (\ref{eq:interm_n_eq}) one can perform the 
integration over 
$d(\Delta{\bf \tilde\Gamma}_r)$ using equation (\ref{eq:delta_fun})
[which reduces to simple replacement of 
$\Delta \tilde h$ by 
$\Delta \tilde h_{sc}(\tilde h,{\bf \tilde e}_r,{\bf \tilde i}_r)$, 
see equation (\ref{eq:delta_fun})].
Also it is  more natural to
 use the {\it reduced} velocity distribution function 
$\tilde \psi_r({\bf \tilde e}_r,{\bf \tilde i}_r)$ defined as
\begin{eqnarray}
\tilde \psi_r({\bf \tilde e}_r,{\bf \tilde i}_r)=
(\mu_1+\mu_2)^{4/3}\psi_r({\bf e}_r)\psi_r({\bf i}_r)=
\frac{1}{4\pi^2\tilde \sigma_{e r}^2\tilde \sigma_{i r}^2}
\exp\left[-\frac{{\bf \tilde e}_r^2}
{2\tilde \sigma_{e r}^2}-\frac{{\bf \tilde i}_r^2}
{2\tilde \sigma_{i r}^2}\right],
\label{eq:psi_defs}
\end{eqnarray}
where 
\begin{eqnarray}
\tilde \sigma_{e,i}=\frac{\sigma_{e,i}}{(\mu_1+\mu_2)^{1/3}}.
\label{eq:tilde_sig}
\end{eqnarray}

Using all these results and definitions
one can finally transform equation (\ref{eq:interm_n_eq}) into
\begin{eqnarray}
&& \frac{\partial N_1(h)}{\partial t}=-2|A|\int\limits_0^{\infty}dm_2
(\mu_1+\mu_2)^{2/3}
\int\limits_{-\infty}^{\infty}d\tilde h |\tilde h|
\Bigg\{N_1(h)N_2
[h-(\mu_1+\mu_2)^{1/3} \tilde h]\nonumber \\
&& -\left.
\int d{\bf \tilde e}_r d{\bf \tilde i}_r 
\tilde \psi_r({\bf \tilde e}_r,{\bf \tilde i}_r)
N_1[h+D(\Delta \tilde h_{sc})]
N_2[h+D(\Delta \tilde h_{sc})-(\mu_1+\mu_2)^{1/3} \tilde h]
\right\}.
\label{eq:n_eq}
\end{eqnarray}
The functions $N_{1,2}(h)$ and $\tilde\psi_r({\bf \tilde e}_r,
{\bf \tilde i}_r)$ are
also functions of time. This equation is analogous to the 
evolution equation (6) of Paper I
but now it can also take into account the dependence of scattering 
properties on the random motion of interacting bodies.


\subsection{Evolution of random velocities.}
\label{subsect:evol_vels}


Let us multiply equation (\ref{eq:general_hill_master}) by ${\bf e}^2
d{\bf e} d{\bf i}$ and, again, integrate over the whole 
vector eccentricity and inclination space. Since integrating the l.h.s. of 
(\ref{eq:general_hill_master}) and the first term in its r.h.s. 
is again trivial
[by definition $\int{\bf e}^2 f_j({\bf e},{\bf i})
d{\bf e} d{\bf i}=2N_j\sigma_e^2$], we concentrate on
 the second term in the r.h.s.  We notice using 
(\ref{eq:f_rel_def}) that the integration of
$\psi_1 \psi_2 {\bf e}^2d{\bf e}$ reduces to
\begin{eqnarray}
\exp\left[-\frac{{\bf e}_r^2}
{2\left(\sigma_{e 1}^2+\sigma_{e 2}^2\right)}\right]~
\int \frac{{\bf e}^2d{\bf e}}{4\pi^2\sigma_{e 1}^2
\sigma_{e 2}^2}\exp\left[-
\frac{\sigma_{e 1}^2+\sigma_{e 2}^2}
{2\sigma_{e 1}^2\sigma_{e 2}^2}
\left({\bf e}-\frac{\mu_2}{\mu_1+\mu_2}
\Delta{\bf e}_r- \frac{\sigma_{e 1}^2 
{\bf e}_r}{\sigma_{e 1}^2+\sigma_{e 2}^2}\right)^2\right]
\nonumber\\
=\psi_r\left({\bf e}_r\right)\left[
2\frac{\sigma_{e 1}^2\sigma_{e 2}^2}
{\sigma_{e 1}^2+\sigma_{e 2}^2}+
\frac{\sigma_{e 1}^4{\bf e}_r^2}
{(\sigma_{e 1}^2+\sigma_{e 2}^2)^2}
+\left(\frac{\mu_2}{\mu_1+\mu_2}\right)^2\left(\Delta {\bf e}_r\right)^2
+2\frac{\mu_2}{\mu_1+\mu_2}\frac{\sigma_{e 1}^2}{\sigma_{e 1}^2+ 
\sigma_{e 2}^2}{\bf e}_r\cdot\Delta {\bf e}_r
\right]
\end{eqnarray}
[Integration over $d{\bf i}$ yields $\psi_r({\bf i}_r)$,
see equations (\ref{eq:f_rel_def})].
Using again the PlDF in the form (\ref{eq:delta_fun}) 
we find that integration of the second part of the r.h.s. of 
(\ref{eq:general_hill_master}) gives
\begin{eqnarray}
&& \int d{\bf \tilde e}_r d{\bf \tilde i}_r 
\tilde\psi_r({\bf \tilde e}_r,{\bf \tilde i}_r) 
N_1[h+D(\Delta \tilde h_{sc})]
N_2[h+D(\Delta \tilde h_{sc})-(\mu_1+\mu_2)^{1/3} \tilde h]
\nonumber\\
&& \times\left[
2\frac{\sigma_{e 1}^2\sigma_{e 2}^2}
{\sigma_{e 1}^2+\sigma_{e 2}^2}+
\frac{\sigma_{e 1}^4{\bf e}_r^2}
{(\sigma_{e 1}^2+\sigma_{e 2}^2)^2}+
\left(\frac{\mu_2}{\mu_1+\mu_2}\right)^2\left(\Delta {\bf e}_{sc}\right)^2
+2\frac{\mu_2}{\mu_1+\mu_2}\frac{\sigma_{e 1}^2}{\sigma_{e 1}^2+ 
\sigma_{e 2}^2}{\bf e}_r\cdot\Delta {\bf e}_{sc}
\right],\label{part1}
\end{eqnarray}
where we have again used function 
$\tilde\psi_r({\bf \tilde e}_r,{\bf \tilde i}_r)$ 
defined in equation (\ref{eq:psi_defs}), and
\begin{eqnarray} 
\Delta {\bf e}_{sc}(\tilde h,{\bf \tilde e}_r,{\bf \tilde i}_r)=
(\mu_1+\mu_2)^{1/3}
\Delta {\bf \tilde e}_{sc}(\tilde h,{\bf \tilde e}_r,{\bf \tilde i}_r)
\label{eq:nonred_de}
\end{eqnarray} 
comes  from equation (\ref{eq:delta_fun}). 

Using these results we can rewrite the evolution equation of
the  r.m.s. eccentricity of planetesimals of mass
$m_1$ in the following form:
\begin{eqnarray}
&& \frac{\partial}{\partial t}\left[2N_1(h)\sigma_{e 1}^2(h)\right]=
-2|A|\int\limits_0^{\infty}dm_2
(\mu_1+\mu_2)^{2/3}
\int\limits_{-\infty}^{\infty}d\tilde h |\tilde h|\nonumber\\
&& \times\Bigg\{2\sigma_{e 1}^2(h)N_1(h)
N_2[h-(\mu_1+\mu_2)^{1/3} \tilde h] -
\int d{\bf \tilde e}_r d{\bf \tilde i}_r \tilde
\psi_r({\bf \tilde e}_r,{\bf \tilde i}_r) N_1(h_1) N_2(h_2)
\nonumber\\
&& \left.\times\left[2\frac{\sigma_{e 1}^2\sigma_{e 2}^2}
{\sigma_{e 1}^2+\sigma_{e 2}^2}+
\frac{\sigma_{e 1}^4{\bf e}_r^2}
{(\sigma_{e 1}^2+\sigma_{e 2}^2)^2}+
\left(\frac{\mu_2}{\mu_1+\mu_2}\right)^2\left(\Delta {\bf e}_{sc}\right)^2
+2\frac{\mu_2}{\mu_1+\mu_2}\frac{\sigma_{e 1}^2}{\sigma_{e 1}^2+ 
\sigma_{e 2}^2}{\bf e}_r\cdot\Delta {\bf e}_{sc}
\right]\right\}.
\label{eq:ecc_eq}
\end{eqnarray}
As usual, it is implied in the last term of this equation that
\begin{eqnarray} 
&& \sigma_{e 1}^2=\sigma_{e 1}^2\left(h_{1}\right),~~~
\sigma_{e 2}^2=\sigma_{e 1}^2\left(h_{2}\right),~~~~~~~
\mbox{where now}    \nonumber \\ 
&& h_{1}=h+D(\Delta \tilde h_{sc}),~~~
h_{2}=h+D(\Delta \tilde h_{sc})-(\mu_1+\mu_2)^{1/3} 
\tilde h, 
\end{eqnarray}
i.e. only the kinematic characteristics of planetesimals of different masses 
at their locations {\it before} the encounter are important.

It should be remembered that the
integral over $d{\bf \tilde e}_r d{\bf \tilde i}_r$ in the r.h.s. of equation
(\ref{eq:ecc_eq}) cannot be easily evaluated in general, 
because $\Delta \tilde h_{sc}$ 
and $\Delta {\bf e}_{sc}$
depend on both 
${\bf \tilde e}_r$ and ${\bf \tilde i}_r$.
However, if the disk kinematic properties and surface density are homogeneous 
(i.e. $\sigma_{e j}$, $\sigma_{i j}$, and $N_j$ are independent of $h$), 
then the integration over
$d{\bf \tilde e}_r d{\bf \tilde i}_r$ can be easily performed
on the first two terms in the square brackets 
of equation (\ref{eq:ecc_eq}) and they vanish.
Planetesimal eccentricities would then evolve only because of the 
presence of the last two terms in square brackets in (\ref{eq:ecc_eq}).
This leads us to the conclusion that these nonvanishing terms 
[third and fourth in the r.h.s. of (\ref{eq:ecc_eq})] are responsible
for the planetesimal {\it self-stirring} ---
kinematic heating in the course of gravitational scattering 
of planetesimals (which is present
even in a completely homogeneous disk). 
The equation of eccentricity evolution in homogeneous disk
then assumes the form [using (\ref{eq:ecc_eq})]
\begin{eqnarray}
\frac{\partial \sigma_{e 1}^2}{\partial t}=
|A|\int\limits_0^{\infty}dm_2
(\mu_1+\mu_2)^{2/3}N_2\left[
\left(\frac{\mu_2}{\mu_1+\mu_2}\right)^2 H_1
+2\frac{\mu_2}{\mu_1+\mu_2}\frac{\sigma_{e 1}^2}{\sigma_{e 1}^2+ 
\sigma_{e 2}^2}H_2\right]
\label{eq:homog_heating}
\end{eqnarray}
where self-stirring coefficients are defined as
\begin{eqnarray}
H_1=\int\limits_{-\infty}^{\infty}d\tilde h |\tilde h|
\int d{\bf \tilde e}_r d{\bf \tilde i}_r \tilde
\psi_r({\bf \tilde e}_r,{\bf \tilde i}_r)\left(\Delta {\bf e}_{sc}\right)^2,
~~~~~H_2=\int\limits_{-\infty}^{\infty}d\tilde h |\tilde h|
\int d{\bf \tilde e}_r d{\bf \tilde i}_r \tilde
\psi_r({\bf \tilde e}_r,{\bf \tilde i}_r)
({\bf e}_r\cdot\Delta {\bf e}_{sc}).
\label{eq:stirring_coeffs}
\end{eqnarray}
Equation (\ref{eq:homog_heating}) and definitions (\ref{eq:stirring_coeffs})
are analogous to equations of evolution of planetesimal eccentricity
dispersion derived by other authors for the case of a homogeneous planetesimal
disk (Ida 1990; Wetherill \& Stewart 1993; SI). 

The other terms in the r.h.s. of equation (\ref{eq:ecc_eq}), 
which disappear in the homogeneous disk, must describe 
the {\it transport} of horizontal energy 
due to gravitational scattering, and are caused by 
(1) the gradients of r.m.s. eccentricity and inclination of planetesimals, 
and (2) the gradients of planetesimal number density. 
To the best of our knowledge, this part of eccentricity 
evolution has never been
taken explicitly into account before. 
In the general case of an inhomogeneous disk (e.g. in the 
presence of a massive embryo which 
can naturally induce gradients of disk properties by its gravity) 
 both contributions (self-heating and transport) are important.

In analogous fashion we can write the equation for the 
inclination evolution:
\begin{eqnarray}
&& \frac{\partial}{\partial t}\left[2N_1(h)\sigma_{i 1}^2(h)\right]=
-2|A|\int\limits_0^{\infty}dm_2
(\mu_1+\mu_2)^{2/3}
\int\limits_{-\infty}^{\infty}d\tilde h |\tilde h|
\nonumber\\
&& \Bigg\{2\sigma_{i 1}^2(h)N_1(h)
N_2[h-(\mu_1+\mu_2)^{1/3} \tilde h]- 
\int d{\bf \tilde e}_r d{\bf \tilde i}_r \tilde
\psi_r({\bf \tilde e}_r,{\bf \tilde i}_r) 
N_1(h_1)
N_2(h_2)
\nonumber\\
&& \left.\times\left[2\frac{\sigma_{i 1}^2\sigma_{i 2}^2}
{\sigma_{i 1}^2+\sigma_{i 2}^2}+
\frac{\sigma_{i 1}^4{\bf i}_r^2}
{(\sigma_{i 1}^2+\sigma_{i 2}^2)^2}+
\left(\frac{\mu_2}{\mu_1+\mu_2}\right)^2\left(\Delta {\bf i}_{sc}\right)^2
+2\frac{\mu_2}{\mu_1+\mu_2}\frac{\sigma_{i 1}^2}{\sigma_{i 1}^2+ 
\sigma_{i 2}^2}{\bf i}_r\cdot\Delta {\bf i}_{sc}
\right]\right\}
\label{eq:inc_eq}
\end{eqnarray}
[the definition of $\Delta {\bf i}_{sc}$
is analogous to definition of $\Delta {\bf e}_{sc}$ in 
equation (\ref{eq:nonred_de})].
We cannot simplify these equations further without making additional 
assumptions about the scattering properties of planetesimals. These 
additional approximations will be made in the next section.


\section{Fokker-Planck expansion.}
\label{sect:FP_expansion}


Significant simplification can be achieved if 
gravitational scattering is weak, i.e. changes in quantities 
characterizing the planetesimal
state ${\bf \Gamma}$ are small compared with the average values of these 
quantities before the encounter. Then a Fokker-Planck type expansion can be
performed (Lifshitz \& Pitaevskii 1981; Binney \& Tremaine 1987). 

This situation is usually the case for planetesimal-planetesimal interactions
because the evolution of the disk's kinematic 
 properties prior to the emergence of embryos
would likely lead to a considerable dynamical ``heating'' of
planetesimal population. This heating brings planetesimal-planetesimal 
interactions into the dispersion-dominated regime
when two interacting planetesimals of types 1 and 2 have 
$\tilde \sigma_{e r}^2+\tilde \sigma_{i r}^2\gg 1$.
Then encounters occur at high relative velocities 
and large-angle scattering is rare. For example, a rocky planetesimal with the
radius $50$ km  at $1$ AU has a corresponding mass of $\approx 10^{21}$ g
and Hill radius $\approx 10^{-4}$ AU. The critical velocity determining
the lower boundary of the dispersion-dominated regime for the 
gravitational interactions between such planetesimals is about
$3$ m~s$^{-1}$, which is likely to be exceeded by the velocity of 
planetesimal epicyclic motion (e.g. see the results of coagulation
simulations by Inaba \etal 2001). Thus, the assumption that
planetesimal-planetesimal scattering occurs in the dispersion-dominated
regime is usually justified in protoplanetary disks.
 
The necessary condition for the Fokker-Planck expansion is that
changes of $h,{\bf e}_r,{\bf i}_r$ are small 
compared with their typical magnitudes. For example, for the 
dimensionless impact parameter $h$
this means that $a\Delta h$ should be much smaller than the 
typical length scale on which the surface density of planetesimals varies.
However, this approximation does not require that the 
relative distance between interacting 
bodies $a h$ must be small in comparison to this typical scale. This
is in contrast to the local approximation that is often used along with 
the Fokker-Planck approach. Indeed, when the 
planetesimal disk is dynamically hot, every 
planetesimal can interact with all others within the reach of its 
own epicyclic excursion. Since eccentricity is large in this velocity
regime the semimajor axis separation of 
interacting planetesimals
$\sim ae$ could be of the order of or bigger than 
$r_H$; then 
local approximation (which would assume that semimajor axis separation 
 of interacting planetesimals is small)
becomes inapplicable. This complication is 
 explicitly taken into account
in our further consideration.


\subsection{Surface density evolution equation.}
\label{subsec:FP:surf_evol}


To implement the Fokker-Planck treatment of the surface density 
evolution equation (\ref{eq:n_eq}),
we first expand the second term of the r.h.s. of equation (\ref{eq:n_eq})
in a Taylor series in $D(\Delta \tilde h_{sc})$ up to the second order. 
Using the notation 
$\alpha=(\mu_1+\mu_2)^{1/3}$ and
$h-(\mu_1+\mu_2)^{1/3} \tilde h=h-\alpha \tilde h$ we obtain that:
\begin{eqnarray}
&& \int d{\bf \tilde e}_r d{\bf \tilde i}_r 
\tilde \psi_r({\bf \tilde e}_r,{\bf \tilde i}_r)
N_1[h+D(\Delta \tilde h_{sc})]
N_2[h+D(\Delta \tilde h_{sc})-(\mu_1+\mu_2)^{1/3} \tilde h]
\nonumber\\
&& =N_1(h)N_2(h-\alpha \tilde h)
-\frac{\mu_2}{(\mu_1+\mu_2)^{2/3}}\frac{\partial}{\partial h}
\left[\langle\Delta \tilde h\rangle
N_1(h)N_2(h-\alpha \tilde h)\right]\nonumber\\
&& +\frac{1}{2}
\frac{\mu_2^2}{(\mu_1+\mu_2)^{4/3}}\frac{\partial^2}{\partial h^2}
\left[\langle(\Delta \tilde h)^2
\rangle
N_1(h)N_2(h-\alpha \tilde h)\right].
\label{eq:part2}
\end{eqnarray}
In obtaining this expression we have taken into account that  
the distribution function $\tilde \psi_r$ is a function 
of $\tilde \sigma_{e r}^2,\tilde \sigma_{i r}^2$,
and, thus, should also be expanded in $D(\Delta \tilde h_{sc})$.
In equation (\ref{eq:part2}) the 
moments of $\Delta \tilde h$ are
\begin{eqnarray}
\langle(\Delta \tilde h)^\beta\rangle=
\int d{\bf \tilde e}_r d{\bf \tilde i}_r\left[\Delta \tilde h_{sc}
(\tilde h,{\bf \tilde e}_r,{\bf \tilde i}_r)\right]^\beta
\tilde \psi_r({\bf \tilde e}_r,{\bf \tilde i}_r)
\label{eq:dh_moments}
\end{eqnarray}
where  $\beta=1,2$, and $\langle(\Delta \tilde h)^\beta\rangle$ 
is a function of $\tilde h, \tilde \sigma_{e r}, \tilde \sigma_{i r}$.
Substituting (\ref{eq:part2}) into (\ref{eq:n_eq}) we obtain
\begin{eqnarray}
\frac{\partial N_1}{\partial t}=
-\frac{\partial}{\partial h}\left(\Upsilon^N_1 N_1\right)
+\frac{\partial^2}{\partial h^2}\left(\Upsilon^N_2 N_1\right)
\label{eq:FP_surf}
\end{eqnarray}
where
\begin{eqnarray}
\Upsilon^N_1(h)=
2|A|\int\limits_0^{\infty}dm_2\mu_2
\int\limits_{-\infty}^{\infty}d\tilde h |\tilde h|
\langle\Delta \tilde h\rangle
N_2(h-\alpha \tilde h),
\label{eq:Ups_1_surf}\\
\Upsilon^N_2(h)=
|A|\int\limits_0^{\infty}dm_2\frac{\mu_2^2}{(\mu_1+\mu_2)^{2/3}}
\int\limits_{-\infty}^{\infty}d\tilde h |\tilde h|
\langle(\Delta \tilde h)^2\rangle
N_2(h-\alpha \tilde h).
\label{eq:Ups_2_surf}
\end{eqnarray}
Thus, instead of the integro-differential equation (\ref{eq:n_eq}) we
have obtained the partial differential equation (\ref{eq:FP_surf}). 

The integration
over $d\tilde h$ in definitions of transport coefficients $\Upsilon^N_1$ and
$\Upsilon^N_2$ takes into account the non-zero 
range of the planetesimal disk over
which a given planetesimal can experience encounters with other bodies
(nonlocal scattering).
In the local approximation, $\langle\Delta \tilde h\rangle$ 
and $\langle(\Delta \tilde h)^2\rangle$ 
would 
fall off rapidly with increasing $\tilde h$; expanding the 
integrand in the expression for $\Upsilon^N_1$ in $\tilde h$
one could easily derive the {\it local} Fokker-Planck evolution 
equation.  This equation 
has already been discussed by Petit \& H\'enon (1987b) and 
in Paper I.


\subsection{Random velocity evolution equation.}
\label{subsec:ecc_evol}


Now we perform the Fokker-Planck expansion 
of equation (\ref{eq:ecc_eq}) for the velocity dispersion.
Again, we need to concentrate on the expansion of the terms under the
$d{\bf \tilde e}_r d{\bf \tilde i}_r$ integral with respect to 
$D(\Delta \tilde h_{sc})$. 

The  Fokker-Planck expansion usually assumes keeping
only the terms of the two leading orders: the first one is linear in the
expansion parameter, but can suffer strong cancellation effects when 
averaging over the scattering outcomes is performed. The second term
 is quadratic
but it does not suffer from 
cancellation during the averaging
and in general has magnitude similar to the first one. 
In our case terms such as $(\Delta \tilde h_{sc})^2{\bf e}\cdot
\Delta {\bf e}_{sc}$ or $\Delta \tilde h_{sc}
(\Delta {\bf e}_{sc})^2$ are third order in the perturbation and 
should be neglected (we will comment on the order of their smallness in \S 
\ref{sect:scattering_coefficients}). But terms like 
$\Delta \tilde h_{sc}{\bf e}\cdot\Delta {\bf e}_{sc}$ or
$(\Delta \tilde h_{sc})^2{\bf e}^2$ should be kept in our expansion.

Performing this procedure as in \S \ref{subsec:FP:surf_evol}, after some
cumbersome but straightforward transformations, we 
can write the Fokker-Planck 
equation for eccentricity evolution in the following form:
\begin{eqnarray}
\frac{\partial}{\partial t}\left[2N_1\sigma_{e 1}^2\right]=
\Upsilon^e_0N_1-\frac{\partial}{\partial h}\left(\Upsilon^e_1 N_1\right)
+\frac{\partial^2}{\partial h^2}\left(\Upsilon^e_2 N_1\right)
\label{eq:FP_ecc}
\end{eqnarray}
where
\begin{eqnarray}
&& \Upsilon^e_0(h)=
2|A|\int\limits_0^{\infty}dm_2(\mu_1+\mu_2)^{4/3}
\int\limits_{-\infty}^{\infty}d\tilde h |\tilde h|
N_2(h-\alpha \tilde h)\nonumber\\
&& \times \left[
\left(\frac{\mu_2}{\mu_1+\mu_2}\right)^2\langle(\Delta {\bf \tilde e}
)^2\rangle
+2\frac{\mu_2}{\mu_1+\mu_2}\frac{\tilde \sigma_{e 1}^2}
{\tilde \sigma_{e 1}^2+ 
\tilde \sigma_{e 2}^2}
\langle{\bf \tilde e}\cdot\Delta {\bf \tilde e}\rangle
\right],
\label{eq:Ups_0_e}\\
&& \Upsilon^e_1(h)=
2|A|\int\limits_0^{\infty}dm_2
\mu_2(\mu_1+\mu_2)^{2/3}
\int\limits_{-\infty}^{\infty}d\tilde h |\tilde h|
N_2(h-\alpha \tilde h)\nonumber\\
&& \times\left[
2\frac{\tilde \sigma_{e 1}^2\tilde \sigma_{e 2}^2}
{\tilde \sigma_{e 1}^2+\tilde \sigma_{e 2}^2}\langle
\Delta \tilde h\rangle+
\frac{\tilde \sigma_{e 1}^4}
{(\tilde \sigma_{e 1}^2+\tilde \sigma_{e 2}^2)^2}
\langle {\bf \tilde e}^2\Delta \tilde h\rangle
+2\frac{\mu_2}{\mu_1+\mu_2}\frac{\tilde \sigma_{e 1}^2}{\tilde \sigma_{e 1}^2+ 
\tilde \sigma_{e 2}^2}\langle({\bf \tilde e}\cdot\Delta {\bf \tilde e})\Delta 
\tilde h\rangle
\right],
\label{eq:Ups_1_e}\\
&& \Upsilon^e_2(h)=
|A|\int\limits_0^{\infty}dm_2
\mu_2^2
\int\limits_{-\infty}^{\infty}d\tilde h |\tilde h|
N_2(h-\alpha \tilde h)\nonumber\\
&& \times\left[
2\frac{\tilde \sigma_{e 1}^2\tilde \sigma_{e 2}^2}
{\tilde \sigma_{e 1}^2+\tilde \sigma_{e 2}^2}\langle
(\Delta \tilde h)^2\rangle+
\frac{\tilde \sigma_{e 1}^4}
{(\tilde \sigma_{e 1}^2+\tilde \sigma_{e 2}^2)^2}
\langle {\bf \tilde e}^2(\Delta \tilde h)^2\rangle
\right].
\label{eq:Ups_2_e}
\end{eqnarray}
The new scattering coefficients  used in these equations are
\begin{eqnarray}
&& \langle\left(\Delta {\bf \tilde e}\right)^2\rangle
=\int \tilde \psi_r({\bf \tilde e}_r,{\bf \tilde i}_r)
\left(\Delta {\bf \tilde e}_{sc}
\right)^2d{\bf \tilde e}_r d{\bf \tilde i}_r,~~~ 
\langle{\bf \tilde e}\cdot\Delta 
{\bf \tilde e}\rangle=\int\tilde  
\psi_r({\bf \tilde e}_r,{\bf \tilde i}_r){\bf \tilde e}_r\cdot\Delta 
{\bf \tilde e}_{sc} d{\bf \tilde e}_r d{\bf \tilde i}_r,\nonumber\\ 
&& \langle{\bf \tilde e}^2
\Delta \tilde h \rangle=\int\tilde  
\psi_r({\bf \tilde e}_r,{\bf \tilde i}_r){\bf \tilde e}_r^2
\Delta \tilde h_{sc} d{\bf \tilde e}_r d{\bf \tilde i}_r,~~~
\langle{\bf \tilde e}^2
(\Delta \tilde h)^2 \rangle=\int\tilde  
\psi_r({\bf \tilde e}_r,{\bf \tilde i}_r){\bf \tilde e}_r^2
(\Delta \tilde h_{sc})^2 d{\bf \tilde e}_r d{\bf \tilde i}_r, 
\label{eq:de_moments}\\ 
&& \langle({\bf \tilde e}\cdot\Delta 
{\bf \tilde e})\Delta \tilde h\rangle=\int\tilde  
\psi_r({\bf \tilde e}_r,{\bf \tilde i}_r)\Delta \tilde h_{sc}
({\bf \tilde e}_r\cdot\Delta 
{\bf \tilde e}_{sc}) d{\bf \tilde e}_r d{\bf \tilde i}_r,\nonumber
\end{eqnarray}
These coefficients, like 
 the old ones ($\langle\Delta \tilde h\rangle, \langle
(\Delta \tilde h)^2\rangle$) are functions of 
$\tilde h, \tilde \sigma_{e 1}^2, 
\tilde \sigma_{e 2}^2, \tilde \sigma_{i 1}^2, \tilde \sigma_{i 2}^2$.
Analytical expressions for these coefficients 
in the two-body approximation will be derived in
\S \ref{sect:scattering_coefficients}.
An analogous evolution equation can be written for $\sigma_{i 1}^2$.

The first term in the r.h.s. of equation (\ref{eq:FP_ecc}) is responsible
for the gravitational stirring of eccentricity, while the last two terms 
describe energy transport and disappear in homogeneous 
planetesimal disks. 
The coefficients in (\ref{eq:FP_surf}) and (\ref{eq:FP_ecc})
 are nonlocal quantities and this is reflected in the 
presence of an integration over $d\tilde h$ in their definitions. 

Equations (\ref{eq:FP_surf})-(\ref{eq:Ups_2_surf}) and 
(\ref{eq:FP_ecc})-(\ref{eq:Ups_2_e}) describe 
self-consistently the evolution of the surface density, eccentricity, 
and inclination at every point
of an inhomogeneous planetesimal disk due to 
planetesimal-planetesimal gravitational
encounters in the dispersion-dominated regime.
The principal approximation is that we follow moments 
rather than whole distribution function.
To close this system we only need to supplement it with the
expressions for various scattering coefficients appearing in evolution 
equations.


\section{Scattering coefficients in the dispersion-dominated regime.}
\label{sect:scattering_coefficients}


In this  section we derive analytical expressions for the scattering
coefficients used in equations (\ref{eq:FP_surf})-(\ref{eq:Ups_2_surf}) and 
(\ref{eq:FP_ecc})-(\ref{eq:Ups_2_e}). In so doing we will always 
assume that 
interaction between planetesimals is in the dispersion-dominated regime,
for the reasons given at the beginning of \S \ref{sect:FP_expansion}.
In the case of embryo-planetesimal scattering it is less clear that the 
scattering is dispersion-dominated since the embryo mass and Hill 
radius are much larger; we consider these additional details in Paper II.

\begin{figure}[t]
\vspace{10.cm}
\includegraphics{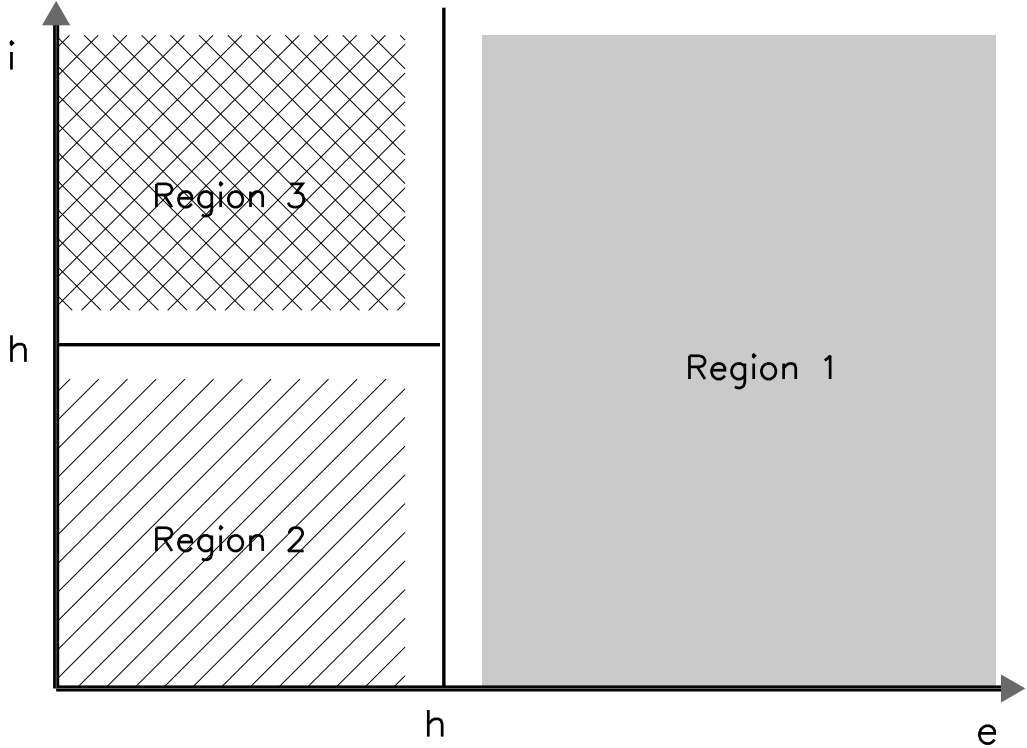}
\caption{Separation of different scattering regimes in the $e-i$ phase
space.}
\label{fig:phase_regions}
\end{figure}

In the dispersion-dominated regime most 
interactions occur between planetesimals separated by $|\tilde h|\gg 1$
(or $|h|\gg r_H$). 
In this case they can be divided
into those that can experience {\it close encounters}, and those which
do not approach each other closely. 
For the planetesimals of the first type 
most of the change of their orbital elements occurs near
 the closest approach point. 
Moreover, this change 
occurs so rapidly that one can consider scattering to be
{\it instantaneous} and {\it local}, i.e. to occur at some point and 
not over an extended part of the trajectory.
These observations considerably 
simplify the analytical treatment of the problem because one only needs to
consider a small region near the encounter point.
Clearly, only planetesimals with 
$\tilde e\ge |\tilde h|$ fall in this category (all the orbital 
parameters are relative and we drop subscript ``r'' in this section).
It will turn out that planetesimals in this part of the 
parameter space ($\tilde e\ge |\tilde h|$ and arbitrary 
$\tilde i$) are the most important contributors to the scattering 
coefficients. We will call this
part of the ${\bf \tilde e}-{\bf \tilde i}$ space ``Region 1'' and
devote \S \ref{subsect:reg1} to its study (see Figure \ref{fig:phase_regions}).

Some planetesimals cannot approach closely --- those with 
$\tilde e < |\tilde h|$. 
This part of the phase space can be split into two more regions:
one with $\tilde i<|\tilde h|$, 
we will call it ``Region 2'' (see Figure \ref{fig:phase_regions}), 
and the other with
$\tilde i>|\tilde h|$ (``Region 3''). 
In the Region 2 planetesimals accumulate 
changes of their orbital 
parameters along a significant portion of their orbits, which 
stretches approximately by several times $\tilde h$ along the $y$-direction. 
In a zeroth approximation, the planetesimal disk in this part of phase space
can be treated as not possessing random motion at all. 
Encounters between planetesimals in Region 2 lead to smaller
change in the orbital elements than encounters in 
Region 1 because close encounters cannot take place in Region 2
(Hasegawa \& Nakazawa 1990; SI).
In Region 3 typical distances at which interaction occurs are
similar to those of Region 2 but relative velocities are higher
(they are magnified by the large value of $\tilde i$), which 
makes scattering in Region 3 weaker than either Region 1 or Region 2.

In general all 3 regions will contribute 
to the scattering coefficients for a given value of $\tilde h$. 
However, their contributions are not of
the same order and their relative magnitudes strongly depend on the value
of $\tilde h$. Assume first that $|\tilde h|\le\tilde \sigma_e$.
Then the distribution  (\ref{eq:Gauss_DF}) 
implies that most of the planetesimals
belong to Region 1 (where $\tilde e\ge|\tilde h|$). Since every individual 
scattering event in Region 1 is also stronger
than those happening in Regions 2 and 3 we can conclude that
 Region 1 strongly dominates the scattering coefficients 
when $|\tilde h|\le\tilde \sigma_e$. However, if 
$\tilde \sigma_e\le|\tilde h|$ the number of planetesimals in Region 1
of the phase space becomes exponentially small, so that 
distant encounters (Region 2) will dominate the scattering coefficients.

For planetesimal-planetesimal scattering we are mostly interested 
in quantities {\it integrated} over some part of planetesimal disk, 
such as the coefficients $\Upsilon^N_j$ and
$\Upsilon^e_j$ in equations (\ref{eq:FP_surf}) and (\ref{eq:FP_ecc}).
They are rather insensitive to the details of the distribution
of the integrands with $\tilde h$. Moreover, one can 
show that the contribution of 
distant encounters (Regions 2 and 3) to the integrated quantities
is still subdominant compared to those from Region 1
(SI). For this reason
we will derive only the contribution to the scattering coefficients 
from Region 1 of the phase space and will completely neglect distant 
encounters when discussing the planetesimal-planetesimal interactions.
The role of distant encounters in the context of 
the embryo-planetesimal scattering is more important and will be 
discussed in Paper II.


\subsection{Scattering coefficients due to close encounters.}
\label{subsect:reg1}


Since our disk evolution equations use $\tilde h, {\bf \tilde e}$, and 
${\bf \tilde i}$ as coordinates and since we are interested in dealing with 
{\it inhomogeneous} planetesimal disks it is natural to use the
Hill formalism 
(Ida \etal 1993; Tanaka \& Ida 1996; SI)
to calculate scattering coefficients. This is opposed to the local velocity
formalism of Hornung \etal (1985) which would be rather awkward for our
purposes.

The treatment of the close encounters in the dispersion-dominated regime
is significantly simplified by the fact that the interaction has
a $3$-dimensional character --- planetesimals can  approach the scatterer 
from all directions. This is in contrast to distant encounters which 
are intrinsically $2$-dimensional. A characteristic feature of 
$3$-dimensional scattering is that if we order planetesimals by their
impact parameter, then equal logarithmic 
intervals in impact parameter contribute equally to the scattering
(Binney \& Tremaine 1987). In the dispersion-dominated regime most 
of this interval corresponds to minimum separations less than $r_H$.
In this case close encounters can be treated by using the {\it two-body}
approximation which neglects the influence of the third body (Sun)
on the process of instantaneous local scattering.

We will derive analytical expressions for the scattering coefficients
(\ref{eq:dh_moments}) and (\ref{eq:de_moments})
averaged over $d{\bf \tilde e}d{\bf \tilde i}$. Switching from
${\bf \tilde e}$ and ${\bf \tilde i}$ back to $\tilde e,\tilde i, \tau,$
and $\omega$ (absolute values of relative eccentricity and inclination and
corresponding constant phases, see \S \ref{subsect:Hill_eq}) we find that
$d{\bf \tilde e}d{\bf \tilde i}=\tilde e d\tilde e~\tilde i d\tilde i ~d\tau
~d\omega$. The phases $\tau/2\pi$ and $\omega/2\pi$
are distributed uniformly in the interval $(0,1)$, while the 
distribution of $\tilde e,\tilde i$ has a form [equation
(\ref{eq:Gauss_DF})]
\begin{equation}
\tilde\psi(\tilde e,\tilde i)d\tilde e d\tilde i=\frac{\tilde e d\tilde e 
~\tilde i d\tilde i}{\tilde \sigma_{e}^2
\tilde \sigma_{i}^2}\exp\left[-\frac{\tilde e^2}{2\tilde \sigma_{e}^2}
-\frac{\tilde i^2}{2\tilde \sigma_{i}^2}\right].
\label{eq:Railey}
\end{equation}

We will perform the averaging in two steps: first we average
over phases $\tau$ and $\omega$ (these averages will be denoted 
by $\langle...\rangle_{\tau,\omega}$), and then we average 
over the absolute values of eccentricity and inclination
(these final averages  will be denoted by simply $\langle...\rangle$). 

Thus, initially we fix $\tilde h,\tilde e,$ and $\tilde i$.
The idea of the calculation is to assume the flux and velocity
of incoming particles near the location of the scatterer
(the planetesimal on which we center the reference frame of relative motion)
to be represented by their unperturbed values
and use them as an input for
the two-body scattering problem. These quantities will be 
given by their values at the scatterer's location ${\bf r}=0$ which means
that we only retain zeroth order terms of their expansion near ${\bf r}=0$. 
Then the components of relative 
planetesimal approach velocity ${\bf v}$ in the $r,\varphi$ and $z$ directions 
are given by evaluating the time derivatives of equations
(\ref{eq:unpert_motion}) at $x=y=z=0$:
\begin{equation}
v_{x}=\pm\Omega r_H \sqrt{\tilde e^2-\tilde h^2},~~~
v_{y}=\Omega r_H \frac{\tilde h}{2},~~~
v_{z}=\pm\Omega r_H \tilde i.
\label{eq:vels}
\end{equation}
Everywhere in this section velocities $v_x,v_y,v_z$
are assumed to be defined relative to the {\it local} circular orbit
(i.e. $v_y=\dot y-2A x$, see \S \ref{subsect:Hill_eq}).

Next we replace the variables $\tau$ and 
$\omega$ by the impact parameter,
$l$, and the angle $\phi$ 
of the orbital plane (see Appendix 8.A of Binney \& Tremaine 1987\footnote{
In the notation of Binney \& Tremaine (1987) $l\equiv b$.}). 
We will use $l$ and $\phi$ as 
an equivalent set of variables
and replace averaging over $\tau/2\pi$ and $\omega/2\pi$ by the averaging
over $l$ and $\phi$. Since we assume that the planetesimal flux is locally 
homogeneous around the scatterer the distribution of planetesimal 
trajectories in $\phi$ is uniform.  

Let us set $\tau_0$ and $\omega_0$ to be the values of $\tau$ and $\omega$
for which the planetesimal orbit passes through the location of the scatterer
${\bf r}=0$. There are $4$ such sets of $\tau$ and $\omega$ [because 
$v_{x}$ and $v_{z}$ have a sign ambiguity in (\ref{eq:vels})]. 
It was demonstrated by Ida \etal (1993) that the 
planetesimal trajectories have impact parameters
smaller than $l$ only when the 
phases $\tau$ and $\omega$ of their orbits 
lie within 4 ellipsoids near 4 sets of $\tau_0$, $\omega_0$ 
in $\tau-\omega$ space. The total surface area covered by these
ellipsoids is given by
\begin{equation}
A(l)=\frac{8}{3}\frac{\pi l^2}{r_H^2}
\frac{v}{\Omega r_H}\frac{1}
{\tilde i|\tilde h|\sqrt{\tilde e^2-\tilde h^2}},
\label{eq:surf}
\end{equation}
where 
\begin{eqnarray}
v=\Omega r_H\sqrt{\tilde e^2+\tilde i^2-(3/4)\tilde h^2}
\label{eq:v_0}
\end{eqnarray}
is the magnitude of the planetesimal approach velocity [see (\ref{eq:vels})].
Thus, particles having impact parameters in the range from $l$ to $l+\Delta l$ 
cover a surface area $(dA/dl) \Delta l$ in $\tau-\omega$ phase space. This
allows us to replace the average of some quantity $f$
with respect to $d\tau d\omega/(4\pi^2)$ by
an integration with respect to $dl d\phi$ using the following
conversion: 
\begin{eqnarray}
\langle f\rangle_{\tau,\omega}=\int\limits_0^{2\pi}\int\limits_0^{2\pi}
f(\tau,\omega)\frac{d\omega d\tau}{4\pi^2}~~
\to~~
\frac{4}{3\pi}\frac{1}{r_H^2}
\frac{v}{\Omega r_H}\frac{1}
{\tilde i|\tilde h|\sqrt{\tilde e^2-\tilde h^2}}
\int\limits_0^{l_{max}}ldl\int\limits_0^{2\pi} f(l,\phi)
\frac{d\phi}{2\pi}.
\label{eq:integrat}
\end{eqnarray}
The upper cutoff distance $l_{max}$ will be specified later.
To be precise, the transition represented by (\ref{eq:integrat}) 
is legitimate only when $l\ll r_H$. Some 
planetesimals pass the scatterer at distances $\ge r_H$ but they 
contribute only weakly to the overall change of the orbital elements
(because of the aforementioned dependence on  {\it logarithmic} and
not linear intervals in $l$). For this reason the errors that arise from
using (\ref{eq:surf}) at
 $|\tau-\tau_0|\sim 1$ or $|\omega-\omega_0|\sim 1$ will hardly 
affect the integrals in (\ref{eq:integrat}).

Using equations (\ref{eq:unpert_motion}), (\ref{eq:reduced_e_i}), 
and (\ref{eq:local_vels}) one finds that 
the various combinations of the 
changes of the relative orbital parameters that we need can
be written as 
\begin{eqnarray}
&& \Delta (\tilde {\bf e}^2)=\frac{1}
{\Omega^2 r_H^2}\left[\Delta (v_x^2)+4\Delta (v_y^2)\right],~~~
\Delta (\tilde {\bf i}^2)=\frac{1}{\Omega^2 r_H^2}\Delta (v_z^2),
\label{eq:incrs_squares}\\
&& \tilde {\bf e}\cdot\Delta \tilde {\bf e}=
\frac{v_x\Delta v_x+4v_y\Delta v_y}{\Omega^2 r_H^2},~~~
\tilde {\bf i}\cdot\Delta \tilde {\bf i}=
\frac{v_z\Delta v_z}{\Omega^2 r_H^2}
\label{eq:incrs_eroducts}
\end{eqnarray}
(the collision is assumed to be instantaneous here). 
Obviously, $\Delta (v_i^2)=2v_i\Delta v_i+(\Delta v_i)^2$. 
From equations (\ref{eq:reduced_e_i}) and (\ref{eq:local_vels})
we can also derive that $\tilde h=(2v_y+\Omega x)/(\Omega r_H)$, meaning that
\begin{eqnarray}
\Delta \tilde h=2\frac{\Delta v_y}{\Omega r_H}~~~\mbox{and}~~~
(\Delta \tilde h)^2=4\frac{(\Delta v_y)^2}{\Omega^2 r_H^2}.
\label{eq:h_v_y}
\end{eqnarray}
We also have
\begin{eqnarray}
(\tilde {\bf e}\cdot\Delta \tilde {\bf e})\Delta \tilde h=
2\frac{v_x\Delta v_x\Delta v_y+4v_y(\Delta v_y)^2}{\Omega^3 r_H^3},~~~
(\tilde {\bf i}\cdot\Delta \tilde {\bf i})\Delta \tilde h=
2\frac{v_z\Delta v_z\Delta v_y}{\Omega^3 r_H^3}.
\label{eq:dh_dei}
\end{eqnarray}

To average these expressions over $\tau, \omega$ we first integrate them
over $d\phi/2\pi$ which means that we need to know 
$\langle v_i\rangle_\phi$ and $\langle v_i v_j\rangle_\phi$
(here $i=x,y,z$ and $\langle ...\rangle_\phi$ 
means averaging over $d\phi/2\pi$).
Two-body scattering in terms of coordinates $l,\phi$ 
was considered in detail elsewhere 
(Binney \& Tremaine 1987) and we will simply use known results
for these averages here:
\begin{eqnarray}
&& \langle \Delta v_i\rangle_\phi=-\Delta v_\parallel \frac{v_i}{v},~~~
\langle \left(\Delta v_i\right)^2\rangle_\phi=
\left(\Delta v_\parallel\right)^2 \frac{v_i^2}{v^2}
+\frac{1}{2}\left(\Delta v_\perp\right)^2 
\frac{v^2-v_i^2}{v^2},\nonumber\\
&& \langle \Delta v_i \Delta v_j\rangle_\phi=\frac{v_i v_j}{v^2}
\left[(\Delta v_\parallel)^2-\frac{(\Delta v_\perp)^2}{2}\right],~~~
i\neq j
\label{eq:deltas}
\end{eqnarray}
where  $v=(v_x^2+v_y^2+v_z^2)^{1/2}$ and
\begin{eqnarray}
&& \Delta v_\perp=\frac{2 l v^3}{G(m_1+m_2)}\left[
1+\frac{l^2 v^4}{G^2(m_1+m_2)^2}\right]^{-1}=2v\frac{l v^2}{\Omega^2 r_H^3}
\left[1+\left(\frac{l v^2}{\Omega^2 r_H^3}\right)^2\right]^{-1},
\nonumber\\
&& \Delta v_\parallel=2 v\left[
1+\frac{l^2 v^4}{G^2(m_1+m_2)^2}\right]^{-1}=2 v\left[
1+\left(\frac{l v^2}{\Omega^2 r_H^3}\right)^2\right]^{-1}.
\label{eq:delta_Vs}
\end{eqnarray}
Here, again,  $l$ is the impact parameter. 

Using these expressions we find  that
\begin{eqnarray}
&& \langle\Delta (\tilde {\bf e}^2)\rangle_\phi=\frac{1}{\Omega^2 r_H^2}\left[
\frac{(\Delta v_\parallel)^2-2v\Delta v_\parallel}{v^2}(v_x^2+4 v_y^2)
+\frac{(\Delta v_\perp)^2}{2v^2}(v_y^2+5v_z^2+4 v_x^2)
\right],\nonumber\\
&& \langle\Delta (\tilde {\bf i}^2)\rangle_\phi=
\frac{1}{\Omega^2 r_H^2}\left[
\frac{(\Delta v_\parallel)^2-2v\Delta v_\parallel}{v^2}v_z^2+
\frac{(\Delta v_\perp)^2}{2v^2}(v_y^2+v_x^2)
\right],\nonumber\\
&& \langle{\bf \tilde e}\cdot\Delta {\bf \tilde e}\rangle_\phi=
-\frac{\Delta v_\parallel}{v}
\frac{v_x^2+4v_y^2}{\Omega^2 r_H^2},\nonumber\\
&& \langle{\bf \tilde i}\cdot\Delta {\bf \tilde i}\rangle_\phi=
-\frac{\Delta v_\parallel}{v}
\frac{v_z^2}{\Omega^2 r_H^2},\nonumber\\
&& \langle\Delta \tilde h\rangle_\phi=
-2\frac{\Delta v_\parallel}{v}
\frac{v_y}{\Omega r_H},\nonumber\\
&& \langle(\Delta \tilde h)^2\rangle_\phi=\frac{4}{\Omega^2 r_H^2}
\left[(\Delta v_\parallel)^2 \frac{v_y^2}{v^2}
+\frac{(\Delta v_\perp)^2}{2} \frac{v_x^2+v_z^2}{v^2}\right],\nonumber\\
&& \langle({\bf \tilde e}\cdot\Delta {\bf \tilde e})
\Delta \tilde h\rangle_\phi=
\frac{2v_y}{\Omega^3 r_H^3}
\left[(\Delta v_\parallel)^2 \frac{v_x^2+4v_y^2}{v^2}
+\frac{(\Delta v_\perp)^2}{2} \frac{3v_x^2+4v_z^2}{v^2}\right],\nonumber\\
&& \langle({\bf \tilde i}\cdot\Delta {\bf \tilde i})
\Delta \tilde h\rangle_\phi=
\frac{2v_y}{\Omega^3 r_H^3}\frac{v_z^2}{v^2}
\left[(\Delta v_\parallel)^2-\frac{(\Delta v_\perp)^2}{2}\right].
\label{eq:i_di_dh0}
\end{eqnarray}

We integrate these expressions over $l$ in the 
range from $0$ to $l_{max}$. In doing this we keep only the leading 
terms in the logarithmic factor which appears when we integrate
$\Delta v_\parallel$ and  $(\Delta v_\perp)^2$ over $l$. In this 
approximation terms proportional to $(\Delta v_\parallel)^2$ vanish.
As a result we find that
\begin{eqnarray}
&& \langle \Delta(\tilde {\bf e}^2)\rangle_{\tau,\omega}
=C(\tilde h,\tilde e,\tilde i)
\left[2\tilde e^2+5\tilde i^2-(15/4)\tilde h^2\right],
\label{eq:finpvs}\\
&& \langle {\bf e}\cdot\Delta {\bf e}\rangle_{\tau,\omega}=-
C(\tilde h,\tilde e,\tilde i)\tilde e^2
\label{eq:finpdf}\\
&& \langle \Delta(\tilde {\bf i}^2)\rangle_{\tau,\omega}
=C(\tilde h,\tilde e,\tilde i)
\left[\tilde e^2-2\tilde i^2-(3/4)\tilde h^2\right],
\label{eq:finqvs}\\
&& \langle {\bf i}\cdot\Delta {\bf i}\rangle_{\tau,\omega}=-
C(\tilde h,\tilde e,\tilde i)\tilde i^2,
\label{eq:finqdf}\\
&& \langle \Delta \tilde h\rangle_{\tau,\omega}=-
C(\tilde h,\tilde e,\tilde i)\tilde h,
\label{eq:del_h}\\
&& \langle (\Delta \tilde h)^2\rangle_{\tau,\omega}=
4C(\tilde h,\tilde e,\tilde i)
\left(\tilde e^2+\tilde i^2-\tilde h^2\right),
\label{eq:del_h2}\\
&& \langle ({\bf e}\cdot\Delta {\bf e})\Delta \tilde h\rangle_{\tau,\omega}=
C(\tilde h,\tilde e,\tilde i)\tilde h
\left(3\tilde e^2+4\tilde i^2-3\tilde h^2\right),
\label{eq:del_h_e_de}\\
&& \langle ({\bf i}\cdot\Delta {\bf i})\Delta \tilde h\rangle_{\tau,\omega}=-
C(\tilde h,\tilde e,\tilde i)\tilde h\tilde i^2,
\label{eq:del_h_i_di}
\end{eqnarray}
where
\begin{eqnarray}
C(\tilde h,\tilde e,\tilde i)= \frac{8}{3\pi}\frac{\ln\Lambda}
{\tilde i|\tilde h|\sqrt{\tilde e^2-\tilde h^2}
\left[\tilde e^2+\tilde i^2-(3/4)\tilde h^2\right]^{3/2}}
\left[1+O\left(\frac{1}{\ln\Lambda}\right)\right],
\label{eq:c_def}
\end{eqnarray}
and Coulomb factor 
\begin{eqnarray}
\Lambda=l_{max}v^2/G(m_1+m_2).
\label{eq:loglam}
\end{eqnarray}
The upper cutoff $l_{max}$ is determined by the disk dimensions and we
set it to be of the order of the disk thickness 
$\tilde i r_H$ (Stewart \& Wetherill 1989; SI). 
Substituting $l_{max}=\tilde i r_H$ and (\ref{eq:v_0}) into 
equation (\ref{eq:loglam}) we find that 
\begin{equation}
\Lambda=\tilde i(c_1 \tilde e^2 + c_2 \tilde i^2)\gg 1,
\label{eq:lambda}
\end{equation}
where $c_{1,2}$ are some constants. By introducing these 
constants we avoid the need
of averaging the logarithmic factors over $\tilde e$ and $\tilde i$;
instead we fix them using numerical data [see Ohtsuki \etal (2002) for
a similar treatment of Coulomb logarithms].
One can see that all the scattering coefficients that we
retain in equations (\ref{eq:FP_surf})-(\ref{eq:Ups_2_surf}) and 
(\ref{eq:FP_ecc})-(\ref{eq:Ups_2_e}) are proportional to $\ln\Lambda\gg 1$.
The coefficients of the third and higher orders in the 
Fokker-Planck expansion do not contain this multiplier. 
Thus our neglect of these higher-order coefficients has a relative 
accuracy of $O\left((\ln\Lambda)^{-1}\right)$.

Using equations (\ref{eq:finpvs})-(\ref{eq:del_h_i_di}) 
one can easily check that
\begin{eqnarray}
&& \langle\Delta(\tilde {\bf e}^2)+\Delta(\tilde {\bf i}^2)-
\frac{3}{4}\Delta(\tilde h^2)
\rangle_{\tau,\omega}=0,
\label{eq:Jac_exact}\\
&& \langle ({\bf \tilde e}\cdot\Delta{\bf \tilde e})\Delta \tilde h+
({\bf \tilde i}\cdot\Delta{\bf \tilde i})\Delta \tilde h-
\frac{3}{4}\tilde h(\Delta \tilde h)^2
\rangle_{\tau,\omega}=0.
\label{eq:Jac_second_order}
\end{eqnarray}
The first equation is implied by the conservation of 
Jacobi constant (\ref{eq:Jacobi}).
The second equation is a statement of 
$\langle\Delta J\Delta \tilde h\rangle_{\tau,\omega}=0$
up to the second order in perturbed quantities [or up to the factors 
$\sim(\ln\Lambda)^{-1}$].

Now we perform the final step in our programme ---
we average the scattering coefficients over the Gaussian distribution of 
relative eccentricity and inclination given by (\ref{eq:Railey}).
To do this we have to perform a series of straightforward but
rather lengthy steps which is described in detail in Appendix 
\ref{app:derivation}. 
Coulomb logarithm $\ln\Lambda$ is always treated as a constant, which 
is justified by its weak dependence on $\tilde e$ and
$\tilde i$. Of course one can do better than this (see Ida \etal 1993)
but for our present purposes such an accuracy is sufficient.

As a result we obtain the following formulae
for our scattering coefficients:
\begin{eqnarray}
&& \langle \Delta(\tilde {\bf e}^2)\rangle
=K(\tilde h,\tilde \sigma_e,\tilde \sigma_i)
\left(-\frac{7}{4}U^0_0+\frac{7}{8}U^1_0-\frac{3}{8}U^1_1\right),
\label{eq:av_evs}\\
&& \langle {\bf \tilde e}\cdot\Delta {\bf \tilde e}\rangle
=-K(\tilde h,\tilde \sigma_e,\tilde \sigma_i)
\left(U^0_0+\frac{1}{8}U^1_0+\frac{1}{8}U^1_1\right),
\label{eq:av_edf}\\
&& \langle \Delta (\tilde {\bf i}^2)\rangle
=K(\tilde h,\tilde \sigma_e,\tilde \sigma_i)
\left(\frac{1}{4}U^0_0-\frac{1}{8}U^1_0+
\frac{3}{8}U^1_1\right),
\label{eq:av_qvs}\\
&& \langle {\bf \tilde i}\cdot\Delta {\bf \tilde i}\rangle
=-K(\tilde h,\tilde \sigma_e,\tilde \sigma_i)
\left(\frac{1}{8}U^1_0-\frac{1}{8}U^1_1\right),
\label{eq:av_qdf}\\
&& \langle \Delta \tilde h\rangle=
-\frac{K(\tilde h,\tilde \sigma_e,\tilde \sigma_i)}{\tilde h}U_0^0,
\label{eq:av_dh}\\
&& \langle (\Delta \tilde h)^2\rangle
=K(\tilde h,\tilde \sigma_e,\tilde \sigma_i)U_0^1,
\label{eq:av_dh2}\\
&& \langle  \tilde {\bf e}^2\Delta \tilde h\rangle=
-\tilde h K(\tilde h,\tilde \sigma_e,\tilde \sigma_i)
\left(U_0^0+\frac{1}{8}U^1_0+\frac{1}{8}U^1_1\right),
\label{eq:e_2dh}\\
&& \langle  \tilde {\bf i}^2\Delta \tilde h\rangle=
-\tilde h K(\tilde h,\tilde \sigma_e,\tilde \sigma_i)
\left(\frac{1}{8}U^1_0-\frac{1}{8}U^1_1\right),
\label{eq:i_2dh}\\
&& \langle  \tilde {\bf e}^2(\Delta \tilde h)^2\rangle=
\tilde h^2 K(\tilde h,\tilde \sigma_e,\tilde \sigma_i)
\left(U_0^1+\frac{1}{8}U_0^2+\frac{1}{8}U_1^2\right),
\label{eq:e_2dh_2}\\
&& \langle  \tilde {\bf i}^2(\Delta \tilde h)^2\rangle=
\tilde h^2 K(\tilde h,\tilde \sigma_e,\tilde \sigma_i)
\left(\frac{1}{8}U_0^2-\frac{1}{8}U_1^2\right),
\label{eq:i_2dh_2}\\
&& \langle ({\bf \tilde e}\cdot\Delta {\bf \tilde e}) \Delta \tilde h\rangle=
\tilde h K(\tilde h,\tilde \sigma_e,\tilde \sigma_i)
\left(\frac{7}{8}U_0^1-\frac{1}{8}U_1^1\right),
\label{eq:e_de_dh}\\
&& \langle ({\bf \tilde i}\cdot\Delta {\bf \tilde i}) 
\Delta \tilde h\rangle=-
\tilde h K(\tilde h,\tilde \sigma_e,\tilde \sigma_i)
\left(\frac{1}{8}U_0^1-\frac{1}{8}U_1^1\right).
\label{eq:i_di_dh}
\end{eqnarray}
In these expressions we use the following notation:
\begin{eqnarray}
K(\tilde h,\tilde \sigma_e,\tilde \sigma_i)=
\frac{4}{3}\frac{\ln\Lambda}
{\tilde \sigma_e^2\tilde \sigma_i^2}e^{-\tilde h^2/(2\tilde\sigma_e^2)},
\label{eq:Ddef}\\
U^\eta_\rho\left(\alpha_e,\alpha_i\right)\equiv
\int\limits_0^\infty dt~ \frac{t^\eta}{(t+1)^{3/2}}
e^{-\frac{1}{2}\left(\alpha_e^2+\alpha_i^2\right)t}
I_\rho\left[\frac{1}{2}\left(\alpha_i^2-\alpha_e^2\right)t\right],
\label{eq:Udef}
\end{eqnarray}
($I_\rho$ is a modified Bessel function of order $\rho$) and it is 
always assumed in equations (\ref{eq:av_evs})-(\ref{eq:e_de_dh}) that
\begin{eqnarray}
U^\eta_\rho=U^\eta_\rho\left(\frac{\tilde h/\tilde \sigma_e}{2\sqrt{2}},
\frac{\tilde h/\tilde \sigma_i}{2\sqrt{2}}\right).
\label{eq:Uused}
\end{eqnarray}
Coefficients $\langle (\Delta \tilde {\bf e})^2\rangle$ and 
$\langle (\Delta \tilde {\bf i})^2\rangle$ can be trivially computed from 
$\langle \Delta(\tilde {\bf e}^2)\rangle$,
$\langle {\bf \tilde e}\cdot\Delta {\bf \tilde e}\rangle$,
$\langle \Delta (\tilde {\bf i}^2)\rangle$, and 
$\langle {\bf \tilde i}\cdot\Delta {\bf \tilde i}\rangle$.
We also use for the factor inside the Coulomb logarithm
the following expression: 
$\Lambda=\tilde \sigma_i(c_1 \tilde \sigma_e^2 + c_2 \tilde \sigma_i^2)\gg 1$. 

We perform some checks and comparisons of our scattering coefficients with
the results obtained by other authors. 
First of all, one can check again in a manner analogous to 
equations (\ref{eq:Jac_exact}) and (\ref{eq:Jac_second_order})
that (\ref{eq:av_evs})-(\ref{eq:i_di_dh})  preserve the Jacobi constant. 
Second we have checked that our formulae for 
$\langle \Delta(\tilde {\bf e}^2)\rangle_{\tau,\omega},
\langle {\bf e}\cdot\Delta {\bf e}\rangle_{\tau,\omega},
\langle \Delta(\tilde {\bf i}^2)\rangle_{\tau,\omega},
\langle {\bf i}\cdot\Delta {\bf i}\rangle_{\tau,\omega}$ and 
$\langle \Delta \tilde h\rangle_{\tau,\omega}$
 agree with  SI and Ida \etal (2000). Third, 
in the case of a homogeneous planetesimal disk integrals of
$\langle\Delta(\tilde {\bf e}^2)\rangle,
\langle\Delta(\tilde {\bf i}^2)\rangle,
\langle {\bf \tilde e}\cdot\Delta {\bf \tilde e}\rangle,$ and 
$\langle {\bf \tilde i}\cdot\Delta {\bf \tilde i}\rangle$
over $(3/2)|\tilde h|d\tilde h$ should reproduce the averaged 
viscous stirring and dynamical friction coefficients 
$\langle P_{VS}\rangle,\langle Q_{VS}\rangle$ and 
$\langle P_{DF}\rangle,\langle Q_{DF}\rangle$ of SI.
We numerically checked that this is really the case.

This completes our calculation of the contribution of Region 1
of ${\bf \tilde e}-{\bf \tilde i}$ space 
($\tilde e>|\tilde h|$, arbitrary $\tilde i$)
to the scattering coefficients in the
dispersion-dominated regime.


\section{Discussion and summary.}
\label{sect:Summary}


We have derived a self-consistent set of equations describing the
coupled evolution of the surface density and kinematic properties 
of a planetesimal disk driven by gravitational encounters between 
planetesimals. The assumption of a dispersion-dominated velocity
regime is used throughout the calculation, which is reasonable
for planetesimal disks in their late evolutionary stages.  Thus this
paper serves as a logical extension of Paper I which was devoted to the 
study of the shear-dominated velocity regime.

The evolution equations (\ref{eq:FP_surf}) and (\ref{eq:FP_ecc}) are of 
advection-diffusion type; the coefficients entering 
them are nonlocal which is in contrast with previous results (Paper I;
Ohtsuki \& Tanaka 2002). This is a natural outcome of the 
scattering in the dispersion-dominated regime since planetesimals can
explore large portions of the disk (compared to their Hill radii) in the
course of their epicyclic motion.

We have also derived analytical expressions for the scattering coefficients
entering the different terms of evolution equations 
(\ref{eq:Ups_1_surf}), (\ref{eq:Ups_2_surf}),
(\ref{eq:Ups_0_e})-(\ref{eq:Ups_2_e}). 
The analytical treatment was enabled by the use of the two-body scattering 
approximation which becomes valid in the dispersion-dominated regime.
Our expressions are accurate up to 
fractional errors $\sim (\ln\Lambda)^{-1}$, where
$\ln\Lambda\gg 1$ is a Coulomb logarithm.  Following the methods developed
in SI and Ohtsuki \etal (2002) for the case of
homogeneous planetesimal disks it might be possible to improve
our calculations by taking subdominant higher order terms into account
(they contribute typically at the level  $\sim 10\%-20\%$ and are neglected
in our present consideration).

Using this system of evolution equations supplemented with the information
about the behavior of the scattering coefficients one can self-consistently 
study the evolution of inhomogeneous planetesimal disks. 
Arbitrary distribution of
masses of interacting planetesimals is allowed for but the evolution of mass
spectrum is not considered in the present study. Thus our equations describe 
the disk evolution on timescales
short compared with the timescale of the mass spectrum evolution. This 
should be a good assumption for studying effects on the disk caused 
by the gravity of massive protoplanetary embryos 
(because they can induce changes of the disk
properties on rather short timescales). This deficiency can also 
be easily remedied in the future when the need to study 
very long-term disk evolution
arises. Our approach can naturally 
incorporate physical mechanisms other than just gravitational scattering, 
for example gas drag or migration [see Tanaka \& Ida (1999)] and we are
going to study their effects in the future. 

In the following paper (Paper II) we describe the embryo-planetesimal 
scattering and derive  equations governing this 
process in various velocity regimes. This would allow us to  provide 
a complete description of the disk evolution caused by both the presence of
isolated massive bodies and the continuous sea of planetesimals.

\acknowledgements

I am indebted to my advisor, Scott Tremaine, for his patient guidance 
and valuable advice which I received during the work on this paper. 
Financial support provided by the Charlotte Elizabeth
Procter Fellowship and NASA grant NAG5-10456
is gratefully acknowledged.


\appendix


\section{Calculation of the instantaneous surface density.}
\label{app:n_n_conversion}


Suppose that we know the behavior of the 
surface density of planetesimal guiding centers $N(h)$ 
in the whole disk as well as the random velocity distribution 
function $\psi(e,i,h)$ 
(we will initially carry out calculations for the case 
of general velocity distribution function). 
We want to compute the 
{\it instantaneous} surface density $N^{inst}(x_0)$ at some
point $x_0$, where $x_0$ is scaled by the reference radius $a_0$ and is thus 
dimensionless.

To perform the desired conversion let us denote the number of
planetesimals per unit $d v_x d v_y d v_z dx_0 dz$ as 
$g^{inst}(v_x,v_y,v_z,x_0,z)$ and 
the same number per unit $de di dh d\tau d\omega$ as 
$g(e,i,h,\tau,\omega)$.  The Jacobian of the 
transformation between these sets of 
coordinates [see equations (\ref{eq:unpert_motion}) and 
(\ref{eq:Hill_to_local})] is
\begin{eqnarray}
J=\frac{a_0^4}{2}\Omega^3 e i
\label{eq:Jacob}
\end{eqnarray}
($z$ and velocities $v_x,v_y,v_z$ are assumed to be dimensional).
Clearly,
\begin{eqnarray}
N^{inst}(x_0)=\int g^{inst}(v_x,v_y,v_z,x_0,z)d v_x d v_y d v_z dz=
\int \frac{2g(e,i,h,\tau,\omega)}{a_0^4\Omega^3 e i}d v_x d v_y d v_z dz.
\label{eq:n_inst0}
\end{eqnarray}
 To compute $N^{inst}$ one needs the following expressions obtained using
(\ref{eq:unpert_motion}) and (\ref{eq:Hill_to_local}):
\begin{eqnarray}
e^2=\frac{v_x^2+4v_y^2}{(\Omega a_0)^2},~~~
i^2=\frac{v_z^2+\Omega^2 z^2}{(\Omega a_0)^2},~~~
h=x_0+\frac{2 v_y}{\Omega a_0}.
\label{eq:translation}
\end{eqnarray}

Now we specify a distribution function 
(\ref{eq:Gauss_DF}) and find that
\begin{eqnarray}
N^{inst}(x_0)=\frac{1}{2\pi^2 a_0^4\Omega^3}
\int N(h)
\frac{d v_x d v_y d v_z dz}{\sigma_e^2(h)\sigma_i^2(h)}
\exp\left[-\frac{1}{\Omega^2 a_0^2}\left(\frac{v_x^2+4v_y^2}{2\sigma_e^2}
+\frac{v_z^2+\Omega^2 z^2}{2\sigma_i^2}\right)\right],
\label{eq:n_inst}
\end{eqnarray}
with $h$ given by (\ref{eq:translation}). We can easily perform the 
integration over $d v_x d v_z dz$.  Switching also from 
$v_y$ to $h$ we finally obtain that
\begin{eqnarray}
N^{inst}(x_0)=\frac{1}{\sqrt{2\pi}}\int
\limits^\infty_{-\infty}
\frac{dh}{\sigma_e(h)}
N(h)\exp\left[-\frac{(x_0-h)^2}{2\sigma_e^2}\right].
\label{eq:N_inst}
\end{eqnarray}

Note that $N^{inst}(x_0)=N(x_0)$ when $\sigma_e\to 0$. Also, if 
$N(h)=const$, then $N^{inst}(x_0)=N$ according to formula (\ref{eq:N_inst}).


\section{Averaging over $\tilde e, ~\tilde i$ distribution.}
\label{app:derivation}


We illustrate the procedure of averaging the scattering coefficients 
over the distribution function (\ref{eq:Railey}), using the coefficient
$\langle{\bf \tilde e}\cdot\Delta {\bf \tilde e}\rangle$ 
as an example.
Calculation of other scattering coefficients is analogous. 
Using equations (\ref{eq:finpdf}) and (\ref{eq:Ddef}) we write:
\begin{eqnarray}
\langle {\bf \tilde e}\cdot\Delta {\bf \tilde e}\rangle=-
\frac{8}{3\pi}\frac{\ln\Lambda}
{|\tilde h|\tilde \sigma_{e}^2\tilde \sigma_{i}^2}
\int\limits_{\tilde h}^\infty \int\limits_0^\infty 
\frac{\tilde e^3 d\tilde e d\tilde i
\exp\left[-\tilde e^2/(2\tilde \sigma_{e}^2)-
\tilde i^2/(2\tilde \sigma_{i}^2)\right]}
{\sqrt{\tilde e^2-\tilde h^2}
\left[\tilde e^2+\tilde i^2-(3/4)
\tilde h^2\right]^{3/2}}.
\label{eq:step1}
\end{eqnarray} 
The lower limit of the integration over $\tilde e$ is set to
$\tilde h$ because only test bodies with eccentricities bigger 
than this value can experience close encounters with the reference
particle. Introducing new variables $x$ and $y$ by
\begin{eqnarray}
x^2=\frac{4}{\tilde h^2}(\tilde e^2-\tilde h^2),~~~
y^2=\frac{4}{\tilde h^2}\tilde i^2,
\end{eqnarray}
we can rewrite the double integral in (\ref{eq:step1}) as
\begin{eqnarray}
\int\int ... =2|\tilde h| e^{-
\tilde h^2/(2\tilde\sigma_{e}^2)}
\int\limits_0^\infty\int\limits_0^\infty
\frac{1+(1/4)x^2}{(1+x^2+y^2)^{3/2}}
e^{-\alpha_e^2 x^2-\alpha_i^2 y^2}dx dy,
\end{eqnarray}
where
\begin{eqnarray}
\alpha_e^2=\frac{1}{8}
\frac{\tilde h^2}{\tilde\sigma_{e}^2},~~~
\alpha_i^2=\frac{1}{8}
\frac{\tilde h^2}{\tilde\sigma_{i}^2}.
\end{eqnarray}
Making another change of variables $x=\sqrt{r}\cos\theta,~
y=\sqrt{r}\sin\theta$ we obtain that
\begin{eqnarray}
&& \int\int ... =|\tilde h| e^{-
\tilde h^2/(2\tilde\sigma_{e}^2)}
\int\limits_0^\infty\frac{dr}{(1+r)^{3/2}}\exp\left[
-\frac{r}{2}(\alpha_e^2+\alpha_i^2)\right]
\nonumber\\  
&& \times
\int\limits_0^{\pi/2}\left(1+\frac{r}{8}+\frac{r\cos 2\theta}{8}
\right)\exp\left[\frac{r\cos 2\theta}{2}(\alpha_i^2-\alpha_e^2)\right]d\theta.
\end{eqnarray}
Using the identity (Gradshteyn \& Ryzhik 1980)
\begin{eqnarray}
\int\limits_0^\pi e^{z\cos t}\cos nt~ dt=\pi I_n(z),
\end{eqnarray}
 where $I_n$ is a modified Bessel function, and
introducing definitions (\ref{eq:Udef}) and (\ref{eq:Uused}) we finally obtain
\begin{eqnarray}
\int\int ... =\frac{\pi}{2}|\tilde h| e^{-
\tilde h^2/(2\tilde\sigma_{e}^2)}
\left(U_0^0+\frac{1}{8}U_0^1+\frac{1}{8}U_1^1\right).
\end{eqnarray} 
Using this result and (\ref{eq:step1})  
we finally arrive at equation (\ref{eq:av_edf}).

\end{document}